\documentclass[%
 reprint,
superscriptaddress,
 amsmath,amssymb,
 aps,
]{revtex4-1}

\newcommand{\neb}{$\bar{\nu}_{e} \; $}

\newcommand{\qeq}{$\theta _{13}$}

\newcommand{\Urfive}{$^{235}$U}
\newcommand{\Ureight}{$^{238}$U}
\newcommand{\Punine}{$^{239}$Pu}
\newcommand{\Puone}{$^{241}$Pu}
\newcommand{\tot}{\theta_{13}}
\newcommand{\stot}{\sin^2 2 \theta_{13}}
\newcommand{\dms}{\Delta m^{2}_{31}}

\newcommand{\anu}{$\bar{\nu}_{e}$ }
\newcommand{\anum}{\bar{\nu}_{e}}

\usepackage{graphicx}
\usepackage{dcolumn}
\usepackage{bm}
\usepackage{amsmath}%
\usepackage{amssymb}%
\usepackage{subfigure}
\usepackage[pagewise,mathlines]{lineno}
\begin{document}
\preprint{APS/123-QED}
\title{Reactor \neb disappearance in the Double Chooz 
experiment}

\newcommand{\Aachen}{III. Physikalisches Institut, RWTH Aachen 
University, 52056 Aachen, Germany}
\newcommand{\Alabama}{Department of Physics and Astronomy, University of 
Alabama, Tuscaloosa, Alabama 35487, USA}
\newcommand{\Argonne}{Argonne National Laboratory, Argonne, Illinois 
60439, USA}
\newcommand{\APC}{APC, AstroParticule et Cosmologie, Universit\'{e} Paris 
Diderot, CNRS/IN2P3, CEA/IRFU, Observatoire de Paris, Sorbonne Paris 
Cit\'{e}, 75205 Paris Cedex 13, France}
\newcommand{\CBPF}{Centro Brasileiro de Pesquisas F\'{i}sicas, Rio de 
Janeiro, RJ, cep 22290-180, Brazil}
\newcommand{\Chicago}{The Enrico Fermi Institute, The University of 
Chicago, Chicago, IL 60637, USA}
\newcommand{\CIEMAT}{Centro de Investigaciones Energ\'{e}ticas, 
Medioambientales y Tecnol\'{o}gicas, CIEMAT, E-28040, Madrid, Spain}
\newcommand{\Columbia}{Columbia University; New York, NY 10027, USA}
\newcommand{\Davis}{University of California, Davis, CA-95616-8677, USA}
\newcommand{\Drexel}{Physics Department, Drexel University, Philadelphia, 
Pennsylvania 19104, USA}
\newcommand{\Hamburg}{Institut f\"{u}r Experimentalphysik, 
Universit\"{a}t Hamburg, 22761 Hamburg, Germany}
\newcommand{\Hiroshima}{Hiroshima Institute of Technology, Hiroshima, 
731-5193, Japan}
\newcommand{\IIT}{Department of Physics, Illinois Institute of 
Technology, Chicago, Illinois 60616, USA}
\newcommand{\INR}{Institute of Nuclear Research of the Russian Academy 
of Science, Russia}
\newcommand{\CEA}{Commissariat \`{a} l'Energie Atomique et aux Energies 
Alternatives, Centre de Saclay, IRFU, 91191 Gif-sur-Yvette, France}
\newcommand{\Livermore}{Lawrence Livermore National Laboratory, 
Livermore, CA 94550, USA}
\newcommand{\Kansas}{Department of Physics, Kansas State University, 
Manhattan, Kansas 66506, USA}
\newcommand{\Kobe}{Department of Physics, Kobe University, Kobe, 
657-8501, Japan}
\newcommand{\Kurchatov}{NRC Kurchatov Institute, 123182 Moscow, Russia}
\newcommand{\MIT}{Massachusetts Institute of Technology; Cambridge, MA 
02139, USA}
\newcommand{\MaxPlanck}{Max-Planck-Institut f\"{u}r Kernphysik, 69117 
Heidelberg, Germany}
\newcommand{\Niigata}{Department of Physics, Niigata University, Niigata, 
950-2181, Japan}
\newcommand{\NotreDame}{University of Notre Dame, Notre Dame, IN 46556-
5670, USA}
\newcommand{\IPHC}{IPHC, Universit\'{e} de Strasbourg, CNRS/IN2P3, F-
67037 Strasbourg, France}
\newcommand{\SUBATECH}{SUBATECH, CNRS/IN2P3, Universit\'{e} de Nantes, 
Ecole des Mines de Nantes, F-44307 Nantes, France}
\newcommand{\Sussex}{Department of Physics and Astronomy, University of 
Sussex, Falmer, Brighton BN1 9QH, United Kingdom}
\newcommand{\Tennessee}{Department of Physics and Astronomy, University 
of Tennessee, Knoxville, Tennessee 37996, USA}
\newcommand{\TohokuUni}{Research Center for Neutrino Science, Tohoku 
University, Sendai 980-8578, Japan}
\newcommand{\TohokuGakuin}{Tohoku Gakuin University, Sendai, 981-3193, 
Japan}
\newcommand{\TokyoInst}{Department of Physics, Tokyo Institute of 
Technology, Tokyo, 152-8551, Japan  }
\newcommand{\TokyoMet}{Department of Physics, Tokyo Metropolitan 
University, Tokyo, 192-0397, Japan}
\newcommand{\Muenchen}{Physik Department, Technische Universit\"{a}t 
M\"{u}nchen, 85747 Garching, Germany}
\newcommand{\Tubingen}{Kepler Center for Astro and Particle Physics, 
Universit\"{a}t T\"{u}bingen, 72076, T\"{u}bingen, Germany}
\newcommand{\UFABC}{Universidade Federal do ABC, UFABC, Sa\~o Paulo, Santo 
Andr\'{e}, SP, Brazil}
\newcommand{\UNICAMP}{Universidade Estadual de Campinas-UNICAMP, 
Campinas, SP, Brazil}
\newcommand{\Aviette}{Laboratoire Neutrino de Champagne Ardenne, domaine 
d'Aviette, 08600 Rancennes, France}
\newcommand{\vtech}{Center for Neutrino Physics, 
Virginia Tech, Blacksburg, VA}
\newcommand{\deceased}{Deceased.}

\affiliation{\Aachen}
\affiliation{\Alabama}
\affiliation{\Argonne}
\affiliation{\APC}
\affiliation{\CBPF}
\affiliation{\Chicago}
\affiliation{\CIEMAT}
\affiliation{\Columbia}
\affiliation{\Davis}
\affiliation{\Drexel}
\affiliation{\Hamburg}
\affiliation{\Hiroshima}
\affiliation{\IIT}
\affiliation{\INR}
\affiliation{\CEA}
\affiliation{\Livermore}
\affiliation{\Kansas}
\affiliation{\Kobe}
\affiliation{\Kurchatov}
\affiliation{\MIT}
\affiliation{\MaxPlanck}
\affiliation{\Niigata}
\affiliation{\NotreDame}
\affiliation{\IPHC}
\affiliation{\SUBATECH}
\affiliation{\Muenchen}
\affiliation{\Tennessee}
\affiliation{\TohokuUni}
\affiliation{\TohokuGakuin}
\affiliation{\TokyoInst}
\affiliation{\TokyoMet}
\affiliation{\Tubingen}
\affiliation{\UFABC}
\affiliation{\UNICAMP}
\affiliation{\vtech}

\author{Y.~Abe}
\affiliation{\TokyoInst}

\author{C.~Aberle}
\affiliation{\MaxPlanck}


\author{J.C.~dos Anjos}
\affiliation{\CBPF}



\author{J.C.~Barriere}
\affiliation{\CEA}


\author{M.~Bergevin}
\affiliation{\Davis}

\author{A.~Bernstein}
\affiliation{\Livermore}

\author{T.J.C.~Bezerra}
\affiliation{\TohokuUni}

\author{L.~Bezrukhov}
\affiliation{\INR}

\author{E.~Blucher}
\affiliation{\Chicago}


\author{N.S.~Bowden}
\affiliation{\Livermore}

% ...

\collaboration{Double Chooz Collaboration}
\date{\today}
\begin{abstract}
The Double Chooz experiment has observed 8,249 candidate 
electron antineutrino events in 227.93 live days with
33.71 GW-ton-years (reactor power $\times$ detector mass
$\times$ livetime) exposure using a 10.3 m$^3$ fiducial volume detector
located at 1050~m from the reactor cores of the Chooz nuclear power plant 
in France. 
The expectation in case of \qeq = 0 is 8,937 events.  The deficit is 
interpreted as evidence of
electron antineutrino disappearance. From a rate plus spectral shape 
analysis we find
$\stot$ = 0.109 $\pm$ 0.030(stat) $\pm$ 0.025(syst).
The data
exclude the no-oscillation hypothesis at 99.8\% CL (2.9$\sigma$).
\end{abstract}
\pacs{Valid PACS appear here}
\maketitle

\section{\label{sec:intro} Introduction}
In the three neutrino paradigm, there are
three mixing angles that can be
measured in neutrino oscillation experiments.
For many years, the CHOOZ reactor neutrino
experiment \cite{bib:Chooz}
had the best limit on the value of \qeq.
Recently, the value of \qeq~ has been shown
to be non-zero by the combination of
fits to KamLAND and solar \cite{bib:global, bib:kamq, bib:g2},
MINOS \cite{bib:minose}, T2K \cite{bib:t2k}
and, more precisely, by the new generation
of reactor antineutrino disappearance
experiments: Double Chooz \cite{bib:dc1},
Daya Bay \cite{bib:db1} and
RENO \cite{bib:reno}.  

The Double Chooz analysis is unique among reactor
experiments in its fit
to the energy spectrum.
In the previous reactor measurements of
\qeq, Double Chooz presented both a rate-only analysis and an analysis 
using both the rate and the
shape of the energy spectrum,
while Daya Bay and RENO presented 
rate-only analyses.  The disappearance of reactor electron
antineutrinos has a well-defined effect on the shape of
that spectrum.  The use of the energy distribution to
constrain the oscillation parameters requires a good understanding
of the energy response of the detector and of the accuracy of
the Monte Carlo.  That understanding is
achieved through multiple calibration techniques, in time, space
and energy.  
\par
This paper 
continues the analysis reported in \cite{bib:dc1} with a larger data set,
a new energy scale definition,
reduced background rates and improved systematic uncertainties.
Additionally, 
the running period has been 
subdivided into a two-reactor-on period
and a one-reactor-on period in the
oscillation fit to help separate signal and background.
\par Reactor antineutrinos are observed using the inverse
beta decay (IBD) reaction 
\mbox{$\bar{\nu}_e + p \rightarrow e^+ + n$}
in which there is
a positron whose signal is
promptly seen, and a neutron, whose delayed signal
is seen after a mean time of about 30 $\mu$s from its
capture in the gadolinium-doped target.
The prompt energy of the positron allows us to determine the
antineutrino energy and observe the antineutrino spectrum.
The energy deposited by the positron
including annihilation is related
to antineutrino energy $E_{\bar{\nu}_e}$ by
$E_{\rm prompt}$ = $E_{\bar{\nu}_e} - T_n - 0.8$ MeV
where $T_n$ denotes the average neutron recoil energy
and is small compared to $E_{\bar{\nu}_e}$.

\par The previous analysis represented 15.34 GW-ton-years
of exposure, taking into account the reactor livetime and the
detector fiducial mass.  Here we re-analyze that data set together
with an additional 18.37 GW-ton-years giving a total of 33.71 GW-ton-years.
In addition the analysis of 22.5 hours of both-reactors-off data allows 
a cross check of 
our estimates of the correlated and accidental backgrounds.

The structure of the paper is as follows.  In Section
\ref{sec:detector} we review the experimental setup
and detector.
Section~\ref{sec:simulations} covers the measurements 
and simulations of the Chooz reactors 
used to predict the unoscillated neutrino spectrum, 
as well as the model 
used to describe the detector.  Event reconstruction including
the energy determination of candidate events is described in
Section~\ref{sec:reconstruction}.
The steps that are used to
identify reactor neutrino candidates are covered in 
Section~\ref{sec:analysis}.
Section~\ref{sec:finalfit} presents the extraction of neutrino
mixing parameters from the measured antineutrino rate and 
energy distribution.
\section{Detector and Method Description}
\label{sec:detector}
\subsection{Overview}
\begin{figure}[b]
  \includegraphics[scale=0.25]{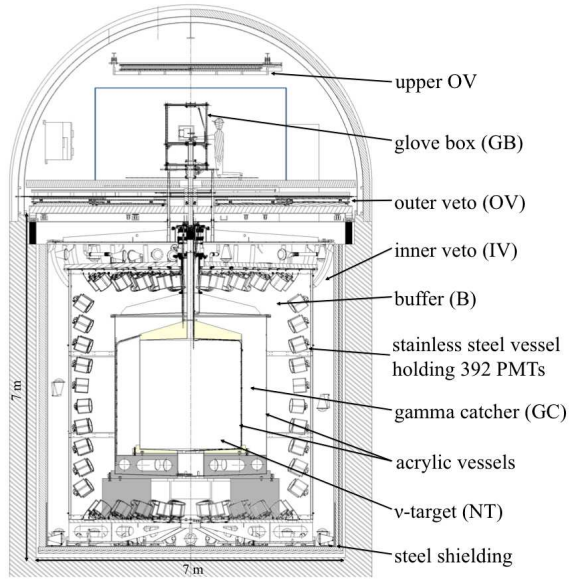}
  \caption{\label{fig:det} A cross-sectional view of the Double Chooz 
detector system.}
\end{figure}

The Double Chooz detector system \cite{bib:dcprop}
consists of a main detector, an outer 
veto, and calibration devices (Figure~\ref{fig:det}).
The main detector is made of four concentric cylindrical tanks 
with a chimney in the center at the top and is 
filled with liquid scintillators or mineral oil.

The innermost $8$~mm thick transparent (UV to visible) acrylic vessel 
contains $10.3$~m$^{3}$ gadolinium loaded liquid scintillator called 
the $\nu$-target (NT).
The NT volume is surrounded 
by the $\gamma$-catcher (GC), a $55$~cm thick Gd-free liquid scintillator layer 
in a second $12$~mm thick acrylic vessel, used to detect gamma rays
escaping from the $\nu$-target. 
Outside the $\gamma$-catcher is the buffer, a 
$105$~cm thick mineral oil layer.
It shields from radioactivity of photomultiplier tubes (PMTs) and 
surrounding rock, and is one of the major improvements over the CHOOZ 
experiment~\cite{bib:Chooz}.
The $390$~10-inch PMTs~\cite{bib:pmt,bib:pmt2,bib:pmt3} are installed on 
the inner wall of the stainless steel buffer tank 
to collect light from the inner volumes.
These three volumes and PMTs constitute central detector system referred
to as the inner detector (ID).
Outside the ID, and optically separated from it by a stainless steel vessel, 
is a $50$~cm thick  inner veto (IV) liquid scintillator.
It is equipped with $78$~8-inch PMTs and functions as a cosmic muon 
veto and as an active shield to spallation 
neutrons produced outside the detector.
The detector is covered and surrounded by $15$~cm of demagnetized steel to 
suppress external gamma rays.
The main detector is covered by an outer veto system (OV) described
in Section~\ref{sec:veto}.
\subsection{Radiopurity}

All parts of the Double Chooz detector have been thoroughly 
screened for their 
content of radioactive isotopes prior to their installation. The 
screening was carried out by direct gamma spectroscopy with a
variety of  
germanium detectors in underground laboratories.
Among them were the large HPGe detector for the non-destructive
radioassay at Saclay \cite{bib:fechner} and the 
GeMPI detector at Gran Sasso \cite{bib:gempi} with a sensitivity of 
about 10\,${\mu\text{Bq}}/{\text{kg}}$
for U and 
Th.
In addition, neutron activation analyses have been performed for 
dedicated parts of the inner detector: the acrylics for NT and GC 
vessels as well as the wavelength shifter PPO \cite{bib:sciprep}. 
The irradiations 
were done at the FRM II research reactor
in Garching, Germany by a thermal neutron flux of 
1.63$\cdot$10$^{13}$\,cm$^{-2}\cdot $s$^{-1}$, with subsequent gamma
spectroscopy in the Garching underground laboratory \cite{bib:hofman}. 

The PMT glass and cavern rock are the main sources of the gamma ray background.
The PMT glass  was 
made from low activity sands using a platinum coated furnace to reduce 
contamination.
Radioactivity of the glass samples was measured during 
development of the low activity glass and production of the 
PMTs \cite{bib:glass}. The average measurements were 13 ppb, 61 ppb and 
3.3 ppb for $^{238}$U, $^{232}$Th and $^{40}$K, respectively assuming 
radio-equilibrium, 
which are much smaller than regular PMT glass.

The design goal of Double Chooz concerning radiopurity is 
no more than $\sim$0.8 
accidental background events per day. 
Along with the radiopurity 
screenings, Double Chooz maintained strict clean-room conditions during 
the setup of the detector with an ISO-level up to 6. 
The analysis 
of BiPo coincidences in the detector data yields concentrations of 
U and Th in NT and GC below the design goal of 10$^{-13}$\,{g}/{g}. 
The accidentals rate is measured to be 
$<$ 0.5 d$^{-1}$, well below our design goal.
The daily rate of correlated background events stemming from 
($\alpha,n$)-reactions of $^{210}$Po on $^{13}$C is estimated 
to be smaller than 0.020 d$^{-1}$ (scaled from the result 
of KamLAND \cite{bib:kamland}),
which is negligibly small compared to the neutrino signal.

\subsection{Double Chooz Liquids}

The CHOOZ experiment was limited in sensitivity by the optical 
instability of its gadolinium-loaded (Gd) scintillator \cite{bib:CHOOZ2}.
Therefore a new type of metal loaded organic liquid scintillator 
was developed for Double Chooz \cite{bib:sciprep}.
The target scintillator used in the NT must fulfill 
the basic requirements of Gd solubility in the solvent of choice, 
optical transparency, radiopurity and chemical stability.
In addition, the organic liquid must be compatible with the detector 
materials in contact with the scintillator, mainly acrylics.
Safety considerations influenced the scintillator design as well.    

Since the rare earth Gd does not dissolve in the required amount 
in the organic solvents used for liquid scintillators,
a metalorganic complex is formed providing higher solubility.
In particular, the complex of choice is a  metal-$\beta$-diketone, 
Gd(thd)$_3$, Gd(III)-tris-(2,2,6,6-tetramethyl-heptane-3,5-dionate).
Such complexes are known for their stability and high vapor pressure.
This allowed us to purify the material by sublimation 
reducing radioimpurities U, Th and K.
The Gd concentration in the NT is 0.123\% by weight, which 
corresponds to about 1~g/liter. 

As scintillator solvent for the NT we have chosen an 
ortho-phenylxylylethane (o-PXE)/n-dodecane mixture at a 
volume ratio of 20/80. To shift the scintillation light into 
a more transparent region, wavelength shifters are added. In 
both scintillators we use PPO (2,5-diphenyloxazole) as primary 
fluor and bis-MSB (4-bis-(2-methylstyryl)benzene) as secondary 
wavelength shifter.

The light yield and density of the 
GC liquid (22.5 m$^3$) were matched simultaneously to the 
NT values \cite{bib:scint}. To achieve this goal, a medicinal white oil was 
added as a third solvent to the GC. The light yield of the GC 
is optimized for homogeneous detector response using Monte Carlo 
simulations. To avoid mechanical stress on the detector vessels 
the densities of all four detector liquids were matched at the 
detector temperature of about 15~$^\circ$C to $0.804~\pm~0.001$ g/cm$^3$. 

\begin{figure}
\includegraphics[scale=0.45]{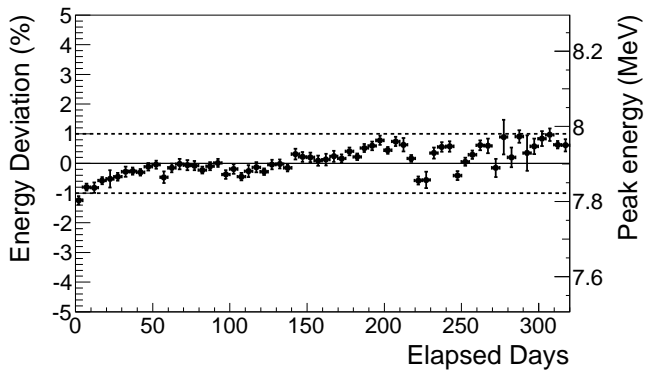}
\caption{
Average target detector response evolution in time, as 
measured by the mean energy of the Gd-capture peak arising from 
interaction of spallation neutrons in the NT.
\label{fig:gstability}}
\end{figure}

The attenuation lengths for wavelengths in
the region of scintillator emission are well above the dimensions of the
corresponding vessels.
Optical stability of the scintillators is demonstrated 
in Figure~\ref{fig:gstability}, 
where the stability of the peak energy of neutron captures on Gd 
is shown. The energy response of the detector was found to be stable 
within 1\% over the data-taking period of about one year. 

The absolute number of H nuclei (``proton number'') as well as
the 
precision on its knowledge are crucial parameters.
The error on the proton number is minimized by using well defined 
and pure chemicals in combination with a precise knowledge 
of the weights of each chemical added in the scintillator production.
The amount of NT scintillator was 
determined after thermalization by a weight measurement with a precision 
of 0.04\%.
The hydrogen fraction in the NT is 13.6\% by weight,
known with 0.3\% relative precision.
This error includes the uncertainties originating from the weights of the
scintillator ingredients. In addition, the error takes into account the
knowledge of the hydrogen content of not fully defined impurities in the
chemicals which are on the per mil level for the main 
components \cite{bib:sciprep}.

A mixture of solvents was used in all detector volumes to allow for
density matching. The 110 m$^3$ of buffer liquid contain a medicinal white
oil (53\% by volume) and an n-alkane mixture (47\% by volume). This liquid was
optimized for transparency and low aromaticity to minimize scintillation
light production in the buffer. The veto volume is filled with 90 m$^3$ of
liquid scintillator, a mixture of linear alkyl benzene (LAB) and
n-alkanes, with 2 g/l PPO as fluor and 20 mg/l bis-MSB as secondary
wavelength shifter.
\subsection{ID Photomultiplier Tubes}
\label{sec:pmt}
The inner detector uses 390 Hamamatsu R7081 10-inch 
PMTs~\cite{bib:hamamatsu} 
to view the target volume.
The glass is a low background type, contributing only a few Hz 
of singles rate in the detector.
The PMTs are operated with a gain of $10^{7}$ at the PMT anode. 
They are submerged in a paraffin oil buffer liquid.
The base circuit is enclosed in a transparent epoxy resin.
Some PMTs are observed to generate light flashes from their base circuit 
through the epoxy resin, causing false triggers.
HV for the 14 worst PMTs was turned off.
Since the signal pattern is different from that of 
the neutrino signal,
the false events are safely removed from the neutrino sample as  
described in Section~\ref{sec:lightnoise}.
The 800 PMTs for both this
and an eventual near detector were characterized 
carefully~\cite{bib:pmt, bib:pmt2, bib:pmt2a}.
The following characteristics 
were measured: for one photoelectron signals, 
the ratio of the one photoelectron peak to the valley between
that peak and the pedestal was 4,
with 1/4 photoelectron thresholds;
the quantum efficiency $\times$ collection efficiency (efficiency 
that photoelectrons produced in the cathode are collected by the first 
dynode) was  ~23\,\%; 
transit time spread was  3~ns (FWHM); 
the afterpulse probability was in average 2.7\,\%; 
the charge output was linear up to 300 photoelectrons per PMT; 
dark hit rate was approximately
2~kHz measured 20 hours after turning on the HV. 
Each PMT is shielded by a mu-metal cylinder to suppress effects from
the gamma shield and the earth's 
magnetic field~\cite{bib:pmt3} and is equipped 
with an angle-adjustable mounting jig.
The PMTs are angled to collect light
more uniformly from the detector.


\subsection{The Inner Veto}

The IV is a cylindrical stainless steel vessel (radius 
3.3\,m and height 6.8\,m) surrounding the ID and optically separated 
by the buffer tank. 
It shields the ID with a 50\,cm thick layer of liquid scintillator 
against external radioactivity and spallation neutrons created by cosmic 
muons. 
At the same time it acts as an active detector identifying cosmic muons 
crossing it. 
The design of the IV was optimized by the use of a MC 
simulation \cite{bib:InnerVetoPaper}, where the emphasis was on a 
high number of detected photoelectrons (PE) per MeV 
deposited in the IV volume and on a high efficiency in rejecting muons 
and correlated background events produced by them. 
The resulting configuration of the IV consists of 78 PMTs, divided into 
three parts: the top has 24 PMTs, the side walls have 12 PMTs at the mid 
way point and the bottom has 42 PMTs.
The 78 8-inch PMTs (Hamamatsu R 1408), which were previously used in the IMB 
and
Super-Kamiokande experiments, 
were tested and modified for use in Double Chooz \cite{bib:ivpmt}. 
Each IV PMT and its base are contained in a stainless steel 
encapsulation, with a transparent PET window at the front end. 
The capsules are filled with mineral oil to match the optical properties 
of the surrounding scintillator. 
All surfaces of the IV are painted with highly reflective white coating
(AR100/CLX coating from MaxPerles \cite{bib:perles}), 
the side walls of the buffer vessel are covered with reflective VM2000 
sheets. 
Using the OV, the muon rejection efficiency was found to be larger than 
$99.99\,\%$ for muons crossing the IV volume. 
\subsection{Electronics and Data Acquisition}
\label{sec:electronics}
\begin{figure}
  \includegraphics[scale=0.30]{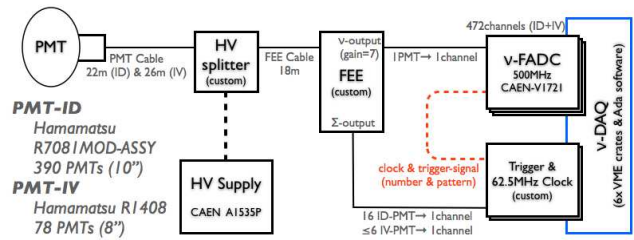}
  \caption{Block diagram of the Double Chooz readout and DAQ systems.
    \label{fig:electronics}}
\end{figure}

%
%
%
%
%
The full readout and data acquisition (DAQ) for both the ID and the IV 
detectors are depicted in Figure~\ref{fig:electronics}.
The functional principle is that digitization of PMT signals (see 
Section~\ref{sec:pmt}) is done by flash-ADC electronics.
As shown in Figure~\ref{fig:electronics}, from left to right, the 
electronics elements are the High Voltage (HV) splitter, the HV supply, 
the Front-End electronics (FEE), the Trigger system~\cite{trigger2} 
and the flash-ADC digitizing 
electronics~\cite{fadc1,fadc2} ($\nu$-FADC).
Each PMT has a single cable for both PMT signal ($5$~mV per PE) and HV 
($\sim 1.3$~kV).
A custom made HV-splitter circuit decouples both components.
The HV is provided by CAEN-A1535P~\cite{bib:caen} supplies.
PMT signals are optimized (amplified, clipped, baseline restored and 
coherent noise filtered) by the FEE for digitization.
The FEE also delivers sum signals, whose amplitude is proportional to 
charge, that are fed into a custom trigger system.
The circuit generating the sum signal subtracts the input amplitude after 
about $100$~ns.
This capability allows the trigger input signals to suffer from minimal 
overshoot that can lead to trigger dead-time.
This same feature works as a high-pass filter: slow signals 
(frequency $\stackrel{<}{\tiny \sim}$ 1~MHz) 
cannot cause a trigger.
The ID PMTs are
separated into two ID super-groups at the trigger level, 
uniformly 
distributed across the volume.
Either super-group can cause a trigger
of the ID based on energy and sub-group multiplicity
information.  The ID triggers at energies about 350~keV.
The trigger efficiency is $100.0$\% above the analysis 
threshold $0.7$~MeV with negligible uncertainty. 
Both energy and sub-group multiplicity information are used to cause IV 
triggers.
The IV triggers at $\sim 10$~MeV which corresponds to 
8 cm of a minimum ionizing muon track.
The $\nu$-FADC system relies on $64$ CAEN-Vx1721(VME64x)~\cite{bib:caen} 
waveform digitizers.
Each card has $8$~channel with $8$-bit flash-ADC (FADC) at $500$~MS/s.
Each channel holds up to $1024$ $4~\mu$s waveforms without readout.
When triggered, the $256$~ns waveform is recorded, containing
$>90$\% of the scintillation light emitted.
Up to $\sim 3$~MeV, a single-PE is deposited per channel, each having 
$\sim 40$~mV amplitude corresponding to around 10 samples per PE.
FADC amplitude saturation leads to some degree of non-linearity for 
$>100$~MeV energies.
Above $500$~MeV, up to a $40$\% non-linearity has been estimated.

The FADC baselines are observed to be stable, showing variations 
below 1 ADC. After power-cycling, small (sub-mV) DC shifts in the 
baseline are observed. Due to under-sampling of the baseline, 
these shifts can cause a bias in the reconstructed charge estimation. This 
bias manifests itself as an effective non-linearity for signals below 
2 PEs and has been thoroughly studied, measured and calibrated 
out, as described in Section
Section~\ref{sec:energy}.

All systems (trigger, ID and IV) are readout by the same DAQ upon any 
trigger of either the ID or IV.
The system is deadtime free, as demonstrated by two monitor systems 
running at $2$~Hz.
The dead time monitor waveforms are, in addition, used to randomly sample 
the detector providing extra baseline monitoring, background and 
dark-current information.
\subsection{The Outer Veto}
\label{sec:veto}
The OV is
installed above the ID, IV and
15 cm of shielding steel.  A lower outer veto is mounted 
directly above the shielding and provides $(x,y)$ coordinate for muons 
passing through 
a 13 m $\times$ 7 m area centered on the chimney; a 110 cm $\times$ 30 cm 
region 
around the chimney is left open.  The lower outer veto has been installed
for 68.9\% of the data presented here, and is used to help reduce
background levels quoted in \cite{bib:dc1}.
An upper outer veto, again measuring 
$(x,y)$ coordinates, has been mounted above the chimney and glove box
used for source insertion, to cover this area.  The upper 
outer veto was not
present for this analysis.

The outer veto is assembled from modules containing 64 scintillator strips,
each 5 cm $\times$ 1 cm 
$\times$  320 cm or 360 cm.
Each strip was extruded with a hole running through its
length, through which a 1.5 mm diameter wavelength-shifting fiber
was threaded. Modules 
are built out of two superimposed 32-strip layers with the top layer
offset
by 2.5 cm from the bottom layer. 
The 64 fibers 
are coupled at one end to a Hamamatsu H8804 multi-anode 
photomultiplier tube (M64); the other fiber ends are mirrored. 
The OV modules are positioned over the inner 
detector in two layers, one with strips oriented in the 
$x$ direction and one in the $y$ direction.
Each M64 is connected to a custom front-end board with a MAROC2 ASIC 
\cite{bib:maroc}
and 
an FPGA. The MAROC2 allows adjustment of the electronic gain of each of 
the 64 channels, which is needed to correct for the 
factor of 2 pixel-to-pixel gain variation in the M64. 
Signals that exceed a common threshold are sent to a 
multiplexed 12-bit ADC, providing charge 
information for hit strips. 
\subsection{Calibration Systems}
\label{sec:calibration}
The experiment is calibrated using light sources, radioactive 
(point-like) sources, and
cosmic rays.

A multi-wavelength LED--fiber system (LI) is used to inject light into 
the inner detector and the inner veto from a set of fixed points.
The optical fibers of the LI
are routed inside the detector and the fiber ends 
are attached to the PMT covers.
Some of the injection points are equipped with diffuser plates to 
widely illuminate the detector.
The other ends of the fibers are connected to blue and UV LEDs 
(385, 425 and 470 nm wavelengths for the ID, and 365 and 475 nm for the IV)
whose flash rate,  light intensity and pulse width
are controlled remotely.
Data are taken with the LI systems regularly.
The LI data are used to measure the PMT and readout electronics
gains and the time offsets and to monitor the stability of those
 gains and
offsets.  

Radio-isotopes $^{137}$Cs, $^{68}$Ge, $^{60}$Co,
and $^{252}$Cf, sealed in miniature capsules, have been deployed in the
NT and GC.
The visible energy response is measured
with a 0.662~MeV gamma (Cs--137), 
2$\times$0.511 MeV annihilation gammas (Ge--68), 
which also corresponds to the
threshold for inverse beta decay, the combination
of 1.173 MeV and 1.333~MeV gammas  (Co--60), 
and the 2.223~MeV gamma from neutron
capture on hydrogen (Cf--252).  
The detector response to
neutrons is calibrated using $^{252}$Cf.
Source rates are at the level of 50 Bq.

Deployments in the NT are realized by lowering the sources from
a glove box at the detector top through the detector chimney.
A motorized pulley-and-weight system, operated from a glove box, 
is used to position sources at positions along the target symmetry axis.
The range of
deployments is from 1 cm above the NT bottom up to the chimney;
the positions of the source are known within 1 mm. 
In the GC, the
source is attached to a motor-driven wire and guided through a rigid hermetic
looped tube (GT). The sources are inserted in the GT near the chimney top.
The loop traverses interior regions of the GC and passes near 
boundaries
with the NT and the buffer. The position of the source along the loop is 
known to 1 cm,
and in the NT boundary region, the perpendicular distance between the source 
and
the target wall is known within 2 mm.   The materials of the source capsules and deployment systems
in the NT and GC are modeled by the detector simulation.

Cosmic rays
are analyzed to identify stopping muons, spallation neutrons, and
cosmogenic radioactive isotopes.   Several thousand spallation
neutrons per day are captured on hydrogen and gadolinium in the 
ID.    

The use of the calibration data for issues of energy
uniformity, stability, non-linearity and absolute calibration is
described in Section~\ref{sec:energy}.  The neutron
detection efficiency from $^{252}$Cf is described 
in Section~\ref{sec:efficiency}.
Good control of uncertainties on detection efficiency
is essential for sensitivity to neutrino 
disappearance with  a single detector.   The detailed 
calibration data allow a precise 
energy-shape fit to the prompt neutrino candidates for the 
most sensitive extraction of \qeq.     


\section{Reactor and Detector Models}
\label{sec:simulations}

\subsection{Thermal Power}
Double Chooz's sources of antineutrinos are the reactor 
cores B1 and B2 at the \'{E}lectricit\'{e} de France (EDF)
Centrale Nucl\'{e}aire de Chooz. Antineutrinos are 
produced 
in nuclear reactors by the $\beta$-decay of the fission products. Four 
main isotopes, \Urfive, \Punine, \Ureight, and \Puone, provide $>$99.7\% 
of 
the fissions and antineutrinos. 
\par
Chooz  B1 and B2 are N4 type pressurized water reactor (PWR) cores, and 
as such are two of the most powerful cores in the world with nominal 
thermal power outputs of 4.25~GW$_{th}$ each. The instantaneous thermal 
power of each reactor core $P_{th}^R$ is provided 
by EDF as a fraction of the 
total power and is evaluated over time steps of $<$1 minute. 
The instantaneous thermal power is 
derived from the in-core instrumentation with the most important variable 
being the temperature of the water in the primary loop.

The in-core instrumentation calibration is tested weekly using the heat 
balance in the secondary loop, which is
heated by the primary loop containing water heated by fissions.
In the secondary loop, steam is generated to 
drive turbines. By using 
 measurements of the heat flow in the secondary loop,
the thermal 
power can be measured. This test is performed with 
the reactor running at full power. The uncertainty at lower power is 
therefore slightly larger.  The 
in-core instrumentation is re-calibrated if it deviates by more than the 
uncertainty in the heat balance measurement.

Since the accuracy of the thermal power measurement determines the 
maximum 
power at which the core can operate, EDF 
 has performed a detailed study of the uncertainty in this 
measurement \cite{EDFTechnicalNote, AFNOR, bib:EDF2011}. The dominant 
uncertainty on the weekly heat balance at the secondary loops comes from 
the measurement of the water flow. At the nominal full power of 4250 MW 
the final uncertainty is 0.5\% (1 $\sigma$ C.L.). Since the amount of data 
taken with one or two cores at intermediate power is small, this 
uncertainty is used for the mean power of both cores. This is smaller 
than the typical uncertainty for PWRs of 
0.7\% \cite{Zelimir2009} and reflects optimizations in the pipe 
geometry of the secondary loop, as well as great care taken to 
understand the sensor uncertainties, including full-scale test 
stands for the most critical sensors.

\subsection{\label{sec:crossperfiss}Mean Cross Section per Fission}

The mean cross section per fission is effectively a spectrum averaged 
cross section. It is given by 
\begin{equation}
\label{eq:crossperfiss}
\langle \sigma_f \rangle = \sum_k \alpha_k \langle \sigma_f \rangle_k 
=  \sum_k \alpha_k \int_{0}^{\infty} dE \, S_{k}(E) \, 
\sigma_{I\!B\!D}(E)
\end{equation}
where $\alpha_{k}$ is the fractional fission rate of the $k^{th}$ 
isotope ($k=$ \Urfive, \Punine, \Ureight, \Puone), $S_{k}(E)$ is the 
reference spectrum of the $k^{th}$ isotope and $\sigma_{I\!B\!D}$ is the 
inverse beta decay cross section. The  determinations of the $\alpha_{k}$ 
require the simulation of 
the reactor core (Section~\ref{sec:frate}). 

The antineutrino spectrum for each fission isotope is the 
result of the beta decays of many different fission products. For 
\Urfive, \Punine, and \Puone, the reference antineutrino spectra are 
derived from measurements of the $\beta$ spectra at the ILL research 
reactor  \cite{SchreckU5,SchreckU5Pu9,SchreckPu9Pu1}. In the case of  
\Ureight, an ab initio calculation of the spectrum is used 
\cite{Mueller2011}. The conversion of the $\beta$ spectra to antineutrino 
spectra has recently been improved by using more data on the 
many $\beta$ transitions and higher order energy 
corrections \cite{Mueller2011,Huber}. We use the conversion scheme 
of \cite{Huber} including corrections for off-equilibrium 
effects\cite{bib:mention}. The uncertainty on these spectra is energy 
dependent but 
is on the order of 3\%.  The new technique for the analysis of the 
$\beta$ spectra has led to an overall change in the normalization 
of the $S_{k}(E)$ that, when applied to previous reactor 
antineutrino experiments, results in measurements that are lower
than predictions 
for experiments at short baselines \cite{bib:mention}.

\subsection{Fission Rate Computation}
\label{sec:frate}
The fractional fission rates $\alpha_k$ of each isotope are needed 
in order to calculate the mean cross section per fission of 
(Equation~\ref{eq:crossperfiss}). They are also required for the 
calculation 
of the mean energy released per fission for reactor $R$: 
\begin{equation}
\label{eq:mpfiss}
\langle E_{f} \rangle_{R} = \sum_{k} \alpha_k  \langle E_{f} \rangle_k.
\end{equation}
The mean energies released per fission per isotope $\langle E_{f} 
\rangle_k$  are summarized in Table \ref{tab:eperfiss}. 
The thermal power one would calculate given a fission is relatively 
insensitive to the specific  fuel composition since the $ \langle E_{f} 
\rangle_k$ differ by $<$6\%; however, the difference in the detected 
number of antineutrinos is amplified by the dependence of the norm and 
mean energy of $S_k$(E) on the fissioning isotope. For this reason, much 
effort has been expended 
in developing simulations of the reactor cores to accurately model 
the evolution of the $\alpha_k$.

Double Chooz has chosen two complementary codes for modeling of the 
reactor cores: MURE and DRAGON 
\cite{MURE,MURE-NEA,DRAGON,bib:jones}. MURE is a 3D full core simulation 
which uses Monte Carlo techniques to model the neutron 
transport in the core. DRAGON 
is a 2D simulation which models the individual fuel assemblies. 
Using some approximations, it
solves the neutron transport equation in the 
core. These 
two codes provide the 
needed flexibility to extract fission rates and their uncertainties.
These codes were benchmarked against 
data from the Takahama-3 reactor and were found to be consistent 
other codes commonly used in the reactor industry
for reactor modeling within the uncertainty 
in the Takahama data \cite{TakahamaPaper}.

The construction of the reactor model requires detailed information 
on the geometry and materials comprising the core. 
The Chooz cores 
are comprised of 205 fuel assemblies. 
For every reactor fuel cycle, approximately one 
year in duration, one third of the assemblies are replaced with 
assemblies containing fresh fuel.  The other two thirds of the 
assemblies are redistributed to obtain a homogeneous neutron flux 
across the core. The Chooz reactor cores contain four assembly 
types that differ mainly in their initial \Urfive ~enrichment. 
These enrichments are 1.8\%, 3.4\% and 4\%.  

The data set presented here
spans fuel cycle 12 for core B2 and cycle 12 
and the beginning of cycle 13 for B1. EDF provides 
Double Chooz with the locations and initial burnup of each 
assembly. 
Based on these maps,
a full core simulation was constructed using MURE for each 
cycle.   In addition, the beginning-of-fuel-cycle composition needs 
to be determined based on the burnup of each assembly. To 
accomplish this, an assembly-level reference simulation is run 
using both MURE and DRAGON for each of the four fuel assembly types. 
The results of the reference simulations are compared to EDF's own 
simulation code APOLLO2-F from which the burnup values are derived. 
The uncertainty due to the simulation technique is evaluated by 
comparing the DRAGON and MURE results for the reference simulation 
leading to a small 0.2\% systematic uncertainty in the fission rate 
fractions $\alpha_k$. 

Once the initial fuel composition of the assemblies is 
known, MURE is used to model the evolution of the full core 
in time steps of 6 to 48 hours, depending on 
the operating conditions of the reactor.
The results from each simulation time step are written to a 
database. This allows the $\alpha_k$'s, and 
therefore the predicted 
antineutrino flux, to be calculated.
The results averaged over the 
current data set are shown in Table \ref{tab:eperfiss}.

The systematic uncertainties on the $\alpha_k$'s are determined 
by varying the inputs and observing their effect on the fission 
rate relative to the nominal simulation. The uncertainties 
considered are those due to the thermal power, boron 
concentration, moderator temperature and density, initial burnup 
error, control rod positions, choice of nuclear databases, choice 
of the energies released per fission, and statistical error of the 
MURE Monte Carlo. The systematic errors associated with each input 
are considered independently and the uncertainties propagated 
quadratically. The correlation coefficients 
among isotopic fission rates 
due to the thermal power constraint 
are also computed, 
and a covariance matrix is constructed with these contributions 
in order to properly account for those
 correlations. 
The uncertainties in the $\alpha_k$'s are listed in 
Table~\ref{tab:eperfiss}. The two largest contributions 
come from the moderator density and control rod positions. 

\begin{table}
\caption{\label{tab:eperfiss} Mean energy released per fission 
 $\langle E_f \rangle _k$ 
from ~\cite{Kopeikin}
and fractional fission rate 
$\langle \alpha_k \rangle $ of the 
isotope k for this data.
}
\begin{center}
\begin{tabular}{c c c}
\hline
Isotope  &  $\langle E_{f} \rangle_k$ (MeV)  & $\langle \alpha_k \rangle$ \\
\hline
\Urfive & 201.92$ \pm $0.46 & 0.496$ \pm $0.016\\
\Punine & 209.99$ \pm $0.60 & 0.351$ \pm $0.013 \\
\Ureight & 205.52$ \pm $0.96 &  0.087$ \pm $0.006 \\
\Puone & 213.60$ \pm $0.65 &0.066$ \pm $0.007 \\
\hline
\end{tabular}
\end{center}
\end{table}

\subsection{Bugey4 Normalization and Antineutrino Rate Calculation}
\label{subsec:nurate}
In the current,
far-only, phase of Double Chooz, the rather large uncertainties 
in the reference spectra of Section \ref{sec:crossperfiss} limited our 
sensitivity to $\theta_{13}$.  To mitigate this effect, the 
normalization of the cross section per fission for each reactor 
is ``anchored" to the Bugey4 rate measurement at 15 m \cite{bib:bugey4}:
\begin{equation}
\label{eq:norm}
\langle \sigma_f \rangle_R=\langle \sigma_f \rangle^{Bugey} + 
\sum_k (\alpha_k^R - \alpha_k^{Bugey})\langle \sigma_f \rangle_k.
\end{equation}
where R stands for each reactor.
The second term corrects for the difference in fuel composition 
between Bugey4 and each of the Chooz cores. This treatment takes 
advantage of the high accuracy of the Bugey4 anchor point (1.4\%) and 
suppresses the dependence on the predicted $\langle \sigma_f \rangle_R$. 
This is due to the smallness of the correction term $(\alpha_{k}^R-
\alpha_k^{Bugey})$. At the same time, the analysis becomes insensitive to 
possible oscillations at shorter baselines due to heavy $\Delta m 
^{2}\sim\, 1~eV^2$ sterile neutrinos.

The expected number of antineutrinos with no oscillation in the 
$i^{th}$ energy bin with the Bugey4 anchor point  becomes:
\begin{eqnarray}
\label{eq:prediction}
\nonumber
N_{i}^{exp,R} = 
	\frac{\epsilon N_{p} }{4\pi} 
	 \frac{1}{L{_R}^2} \frac{P_{th}^R}{\langle E_{f} \rangle_R } 
\\
\times
	 \left(
	 \frac{ \langle \sigma_f \rangle_R}{ \left( \sum_{k} \alpha_{k}^{R} 
\langle \sigma_{f} \rangle_{k} \right)}
	 \sum_{k} \alpha_k^R \langle \sigma_f \rangle_k^i
	 \right)
\end{eqnarray}
where $\epsilon$ is the detection efficiency,  $N_{p}$ is the number of 
protons in the target, $L{_R}$ is the distance to the center of each 
reactor, and $P_{th}^R$ is the thermal power. The variable $\langle E_{f} 
\rangle_R$ is the mean energy released per fission defined in Equation 
\ref{eq:mpfiss}, while $\langle \sigma_f \rangle_R$  is the mean cross 
section per fission defined in Equation \ref{eq:norm}.  The three 
variables $P_{th}^R$,   $\langle E_{f} \rangle_R$ and $\langle \sigma_f 
\rangle_R$ are time dependent with  $\langle E_{f}  \rangle_R$ and 
$\langle \sigma_f \rangle_R$ depending on the evolution of the fuel 
composition in the reactor and $P_{th}^R$ depending on the operation of 
the reactor.

A covariance matrix $M_{ij}^{exp} =
\delta N_i^{exp}\delta N_j^{exp}$ is constructed using the 
uncertainties listed in Table \ref{tab:experror}.  This matrix 
is constructed in terms of real energy and is converted into 
reconstructed energy by running multiple simulations drawn from 
a Cholesky decomposition of $M_{ij}^{exp}$. 
For these simulations, the full detector Monte Carlo 
described below is used. The use of 
Equation~\ref{eq:prediction} to construct the covariance matrix 
allows time and spectral information to propagate to the 
final analysis.

\begin{table}
\caption{\label{tab:experror} The uncertainties in the 
antineutrino prediction. All uncertainties are assumed to be 
correlated between the two reactor cores. They are assumed to 
be normalization and energy  (rate and shape) unless noted as 
normalization only.  }
\begin{center}
\begin{tabular}{l l l}
\hline
Source   &  Normalization Only ~~~~& Uncertainty [\%]\\
\hline
$P_{th}$ & yes & 0.5\\
$\langle \sigma_f \rangle^{Bugey}$ & yes  & 1.4 \\
$ S_{k}(E)\sigma_{I\!B\!D}(E_{\nu}^{true}) $ & no  & 0.2 \\
 $\langle E_{f} \rangle$ & no & 0.2 \\
$L{_R}$ & yes & $<$0.1 \\
$\alpha_k^R$ & no & 0.9 \\
\hline
Total & & 1.8 \\
\hline
\end{tabular}
\end{center}
\end{table}

The IBD cross section used is the simplified form 
from Vogel and Beacom \cite{Vogel1999} :
\begin{equation}
\sigma_{I\!B\!D}(E_{\nu}^{true}) =   E_{e^{+}}K \sqrt{E_{e^{+}}^2 - 
m_{e}^2} 
\end{equation}
where
\begin{equation}
E_{e^{+}}  = \frac{1}{2} \left( \sqrt{m_{n}^2 - 4 m_{p} \left( -
E_{\nu} + \Delta + \frac{\Delta^2 - m_e^2}{2m_p} \right)}-
m_n \right)
\end{equation}
and $m_e$ and $E_{e^+}$ are the positron mass and energy. The variables 
$m_n$ and $m_p$ are the masses of the neutron and proton with $\Delta=m_n 
- m_p$. The constant $K$ is inversely 
proportional to the neutron lifetime. We use the MAMBO-II measurement 
of the neutron lifetime \cite{MAMBOII} and find 
$K=0.961\times10^{-43}$ cm$^{2}$ MeV$^{-2}$. 

\subsection{Detector Model}
\label{sec:monte}

We model the detector response using a detailed Geant4 \cite{Ref:Geant4} 
simulation with
enhancements to the scintillation process, photocathode optical surface
model, and thermal neutron model. Apart from these additions, the physics
list is similar to Geant4's \verb+QGSP_BERT_HP+ reference physics 
list \cite{bib:lists},
without processes for high-mass hadrons.
Our custom scintillation process implements detailed light waveforms, spectra,
re-emission, and Birks-law \cite{bib:birks}
quenching.  Our photocathode model is based on
a standard mathematical model of a thin, semitransparent surface with
absorption and refractive index \cite{Ref:Motta.NIM.A539.217.2005}, and
also includes the collection efficiency for photoelectrons
as a function of position of emission on the photocathode.
Our custom neutron thermalization process implements molecular
elastic scattering for neutrons under 4~eV and a radiative
capture model with improved final state gamma modeling.

The simulation models the detector geometry to a fine level of detail,
particularly with regard to the geometry of the phototubes and
mu-metal shields and of all materials near the active volume such as
tank walls and supports. The orientation and positions of the
phototube assemblies were set using data from a photographic survey
with sub-mm accuracy. The dimensions of the tank walls and supports
were checked by experimenters during assembly and installation, and
placement also verified by photographic survey.

Simulated IBD events are generated with run-by-run
correspondence of MC to data, with fluxes and rates calculated as described
in Section~\ref{subsec:nurate}.
Radioactive decays in calibration sources
and spallation products were simulated using detailed models of nuclear
levels, taking into account branching ratios and correct spectra for
transitions \cite{bib:branching}.

Optical parameters used in the detector model are based on detailed
measurements made by the collaboration.  The relative light yield of the
NT compared to the GC was measured using a Compton
backscatter peak method in order to select scattered electrons with fixed
energy \cite{bib:scint}. Tuning of the absolute and
relative light yield in the simulation was done with calibration data.
The scintillator emission spectrum was measured using a Cary
Eclipse fluorometer \cite{bib:cary}. 
The photon emission time
probabilities used in the simulation are obtained with a dedicated
laboratory setup \cite{bib:sciprep}. For the ionization
quenching treatment in our MC, the light output of the scintillators after
excitation by electrons \cite{Ref:Aberle.JINST.6.P11006.2011} and alpha particles \cite{bib:aberlep}
of different energies was
measured.  
The non-linearity in light
production in the simulation has been adjusted to match these data.
The attenuation and re-emission probabilities of each of the scintillator
components in the relevant wavelength range are implemented in the
MC.  
The fine-tuning of the total attenuation was made 
using measurements of the complete scintillators
\cite{bib:sciprep}.
Other measured optical properties include reflectivities of various
detector surfaces and indices of refraction of detector materials.

\subsection{Readout System Simulation}
\label{sec:ross}
%
%
%
%
%
The Readout System Simulation (RoSS) accounts for the response of 
elements associated with detector readout, such as from the PMTs, 
FEE, FADCs, trigger system and DAQ.
The simulation relies on the measured probability distribution function 
(PDF) to empirically characterize the response to each single PE as 
measured by the full readout channel.
The Geant4-based simulation calculates the time at which each PE strikes the 
photocathode of each PMT.
RoSS converts this time-per-PE into an equivalent waveform as digitized 
by FADCs.
A dedicated setup was built to measure most of the necessary 
PDFs as well as to tune the design of the full readout 
chain.
Channel-to-channel variations, such as gains, baselines, noise, 
single PE 
widths, etc., are taken into consideration, to
accurately predict
dispersion effects.
This capability allows the simulation to exhibit 
non-linearity effects as observed in the data, as described in 
Section~\ref{sec:electronics}.
After calibration, the MC and data energies agree 
within $1$\%.
About $25$\% of the width of the calibrated H-capture (2.2~MeV 
$\gamma$ line) results from readout effects;  i.e., effects beyond 
photon-statistics fluctuations.

\subsection{Monte Carlo \anu Event Generation} \label{sec:nuEventGen}

A set of Monte Carlo \anu events representing the expected signal
for the duration of physics data-taking is created based on the
formalism of Equation~\ref{eq:prediction}.  
The calculated IBD rate is used
to determine the rate of interactions.  Parent fuel nuclide and
neutrino energies are sampled from the calculated neutrino production
ratios and corresponding spectra, yielding a
properly normalized set of IBD-progenitor neutrinos.

Once generated, each event-progenitor neutrino is assigned a random
creation point within the originating reactor core.  The event is
assigned a weighted-random interaction point within the detector
based on proton density maps of the detector materials.  
In the 
center-of-mass frame of the $\nu-p$ interaction, a 
random positron direction is
chosen, with the positron and neutron of the IBD event
given appropriate momenta based on the neutrino energy and decay
kinematics.  These kinematic values are then boosted into the
laboratory frame.
The resulting positron and neutron momenta and originating vertex are
then available as inputs to the Geant4 detector simulation.  ``Truth"
information regarding the neutrino origin, baseline, and energy are
propagated along with the event, for use later in the oscillation
analysis.



\section{Reconstruction}
\label{sec:reconstruction}
\subsection{Pulse Reconstruction}
\label{sec:pulse}



The pulse reconstruction provides the signal charge and time
in each PMT.  
Pulser triggers are taken with a rate of 1 Hz in order to provide 
accurate information about the baseline for each of the 468 readout 
channels. 
The baseline mean ($B_{\rm mean}$) and rms ($B_{rms}$)
are computed using the full readout window (256 ns).   

The integrated charge ($q$) is defined as the sum of digital
counts in each waveform sample over the integration window, once 
the pedestal has been
subtracted.
The pedestal is computed as the integration of $B_{\rm mean}$ 
over the same window. 
In order to improve the charge resolution, 
the size of the integral window has been set to a
112 ns subsample of the readout one, based on the width of the  
single PE signals.

In order to find the pulses within the readout window, a dynamic window 
algorithm is used. The algorithm searches for the 112 ns window which 
maximizes the integral.
In the absence of an actual PE signal, this algorithm would 
reconstruct the largest noise fluctuation, leading to a bias in the 
charge reconstruction. 
To address this,  we introduce two requirements: $\ge$ 2 ADC 
counts in the maximum bin, and 
$q > B_{rms}\times\sqrt{N_{s}}$, where $N_{s}$ is the number of 
integrated waveform samples 
(56 for a 112 ns window). 
For each pulse reconstructed,  the start time is computed 
as the time when the pulse reaches
20\% of its maximum.  This time is then corrected by the 
PMT-to-PMT offsets obtained with the LI system.
\subsection{Vertex Reconstruction}
\label{sec:vertex}
Vertex reconstruction in Double Chooz is not used for
event selection, but is used for event energy reconstruction.
It is based on a maximum charge and 
time
likelihood algorithm which utilizes all hit and no-hit information in the
detector.
Assuming the event to be a point-like source of light characterized by 
the set
\begin{equation} {\bf X} = (x_0,y_0,z_0,t_0,\Phi) \end{equation}
where $(x_0,y_0,z_0)$ is the event position in the detector, $t_0$ is the 
event
time and $\Phi$ is the light intensity per unit solid angle
(expressed in photons/sr), the 
amount of
light and prompt arrival time at the $i$-th PMT can be predicted as
\begin{equation} \mu_i = \Phi\,\epsilon_i\,\Omega_i\,A_i \end{equation}
and
\begin{equation} t_i^{(pred)} = t_0 + \frac{r_i}{c_n} \end{equation}
respectively, where $\epsilon_i$ is the quantum efficiency of the PMT,
$\Omega_i$ is the solid angle subtended by the PMT at a distance $r_i$ 
from
the event vertex, $A_i$ is the light transmission amplitude, and $c_n$ is 
the
effective speed of light in the medium.

The event likelihood is defined as
\begin{equation} {\cal L}({\bf X}) = \prod_{q_i=0}f_q(0;\mu_i)
\prod_{q_i>0}f_q(q_i;\mu_i)f_t(t_i;t_i^{(pred)},\mu_i)
\end{equation}
where the first product goes over the PMTs that have not been hit, while 
the
second product goes over the remaining PMTs that have been hit (i.e., 
have
a non-zero recorded charge $q_i$ at the registered time $t_i$).
$f_q(q_i;\mu_i)$ is the probability to measure a charge $q_i$
 given an expected charge $\mu_i$, and $f_t(t_i;t_i^{(pred)},\mu_i)$ is the probability to measure
 a time $t_i$ given a prompt arrival time $t_i^{pred}$ and predicted charge
 $\mu_i$.
These are obtained from MC simulations and verified against the physics 
and
calibration data.
The task of the event reconstruction is to find the best possible set of
event parameters ${\bf X}_{min}$ which maximizes the event likelihood
${\cal L}({\bf X})$, or equivalently, minimizes the negative 
log-likelihood function
\begin{eqnarray}
\nonumber
 F({\bf X}) = -\ln {\cal L}({\bf X})
              = - \sum_i       \ln f_q(q_i;{\bf X})\\
                - \sum_{q_i>0} \ln f_t(t_i;{\bf X})
              = F_q({\bf X}) + F_t({\bf X}). 
\end{eqnarray}
Note that the event reconstruction can be performed using either
one or both of the 
two
terms in the expression above, $F_q$ for a charge-only reconstruction, or
$F_t$ for a time-only reconstruction; utilizing both components enhances 
the
accuracy and stability of the algorithm.

The performance of the Double Chooz reconstruction has been evaluated 
in situ using radioactive sources deployed at known positions along 
the $z$-axis in
the target volume, and off-axis in the guide tubes.
The sources are reconstructed with a spatial resolution of 32 cm for
$^{137}Cs$, 24 cm for $^{60}Co$, and 22 cm for $^{68}Ge$.
%
\subsection{Muon tagging and reconstruction}
\label{sec:muon}
Cosmic muons passing through the detector or the nearby rock induce
backgrounds which are discussed in the next section.
A through-going (stopping) muon typically deposits
$160$~MeV ($80$~MeV) in the IV which triggers
above about $10$~MeV.  The IV trigger rate is $46$ s$^{-1}$.
All muons in the ID are tagged by the IV except some stopping muons which
enter the chimney.
Muons which stop in the ID and their resulting Michel $e$ can be
identified by demanding a large 
energy deposition (roughly a few tens of MeV) in the ID.
An event is tagged as a muon if there is $>5$~MeV in the
IV or $>30$~MeV in the ID.

Several tracking algorithms have been developed to 
reconstruct these muons. 
IV reconstruction is based on a maximum likelihood algorithm 
utilizing the arrival times of the earliest photons to hit each PMT,
while ID reconstruction utilizes the spatial pattern of hit times.  
The forward wavefront of scintillation light from a relativistic 
track propagates at the Cerenkov angle, thus allowing the same 
algorithm to be used for tracks in the NT, GC, and non-scintillating buffer.
Using MC and the OV as reference, the lateral resolution at 
the detector center has been determined to be 35 cm for ID and 
60 cm for IV muons.

\subsection{Light Noise Rejection}
\label{sec:lightnoise}
The background known as {\em light noise} is caused by a sporadic 
spontaneous flashes of some PMT bases. The characteristic signature is light 
mainly localized to one PMT base and spread out in time among the other 
PMTs after many reflections from the detector surfaces. 
This background can be discriminated from physics events based on the fact that
the detected light is spread less homogeneously across the
detector for light noise events.
Light noise 
is rejected by demanding both a small value of $Q_{\rm max}$/$Q_{\rm tot}$, where 
$Q_{\rm max}$ is the maximum charge recorded by a single PMT and $Q_{\rm tot}$ is 
the total ID charge collected in a trigger, and large
values of rms($t_{\rm start}$), which is the standard deviation of the 
distribution of the start time ($t_{\rm start}$) of the first 
pulse on each PMT.

%
\subsection{Energy Reconstruction}
\label{sec:energy}
The visible energy ($E_{vis}$)  
provides the absolute calorimetric estimation of the energy 
deposited per trigger.
$E_{vis}$ is a function of the calibrated $P\!E$ (total number of 
photoelectrons):
\begin{equation}
E_{vis} 
= P\!E^m(\rho,z,t)
\times f_u^m(\rho,z)
\times f_s^m(t)
\times f_{MeV}^m
\label{eq:evis},
\end{equation}
\noindent where
$P\!E = \sum_{i}\, pe_{i} =
\sum_{i}\, q_{i} / 
gain_{i}( q_{i})$.
Coordinates in the detector are $\rho$ and $z$, $t$ is time,
$m$ refers to data or Monte Carlo (MC) and $i$ refers to each good channel.
The correction factors $f_u$, $f_s$ and $f_{MeV}$ correspond, 
respectively, to the spatial uniformity, time stability and P\!E/MeV
calibrations.
Four stages of calibration are carried out to render $E_{vis}$ linear, 
independent of time and position, and consistent between data and MC.
Both the MC and data are subjected to the same stages of calibration.
%
%
%
%

\begin{figure}
  \includegraphics[scale=0.4]{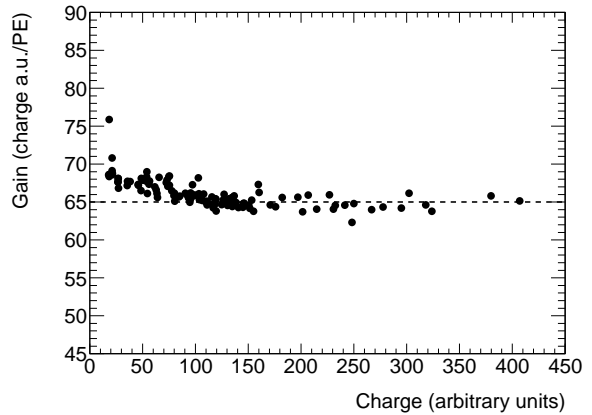}
  \caption{
Demonstration of the linear PE calibration for one channel. 
The gain versus charge is shown. The dashed line highlights 
the constant component (linear behavior) of the gain 
observed at large charges. The calibration parametrizes 
this curve to correct the non-linear component (deviation 
from constant) of the gain, making the PE corrected energy 
scale linear to within 2\%.
    \label{fig:es-nlpe}}
\end{figure}

The sum over all good channels of the reconstructed raw charge ($q_{i}$, 
see Section~\ref{sec:pulse}) from the digitized waveforms is the basis of 
the energy estimation.
Good channels are those identified and tagged as well behaved 
by fast online analysis based on waveform information.
Only a very few channels are sporadically
not good and are, thus, excluded from the 
calorimetric estimation.
The limited sampling of the waveform baseline 
estimation can be biased~\cite{fadc1} leading to a non-linearity at about 
1~$P\!E$ charge equivalent.
Figure~\ref{fig:es-nlpe} shows the effect for a representative channel.
A similar curve is used to calibrate the MC.
The $P\!E$ calibrated charge ($pe_{i}$) is defined as $pe_{i} = q_{i} /
gain_{i}(q_{i})$.
One $gain_{i}(q_{i})$ curve is generated upon each power-cycle 
episode.
%
%
Due to the average light level ($\sim 230~P\!E$/MeV), the non-linear bias 
of the single-$P\!E$ calibration can have up to a $10$\% effect for energies below 
$3$~MeV, if not corrected.  
%
%
%
%

\begin{figure}
\includegraphics[scale=0.4]{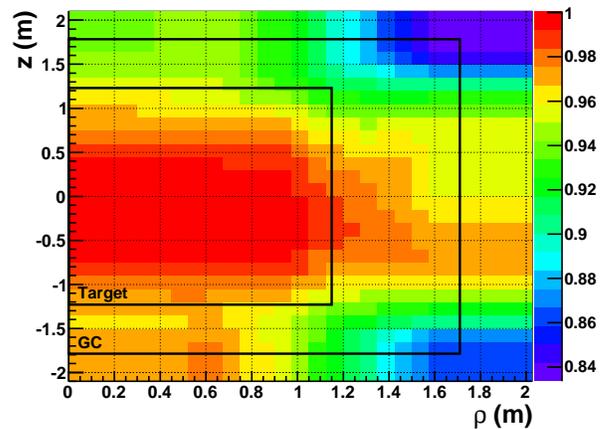}
  \caption{
    Detector calibration map, in cylindrical coordinates ($\rho$,$z$), as 
sampled with spallation neutrons capturing in H across the ID.
    Response variations are quantified as the fractional response with 
respect to the detector center.
    Largest deviation in NT are up to $5$\%.
    A similar map is constructed with MC for calibration of its slightly 
different response uniformity pattern.
    \label{fig:map}}
\end{figure}

The $P\!E$ response is position dependent for both MC and data.
Calibration maps were created such that any $P\!E$ response for any event 
located at any position ($\rho$,$z$) can be converted into its 
response as if measured at the center 
of the detector ($\rho=0,z=0$): 
$  P\!E^m_{\odot}  =   P\!E^m(\rho,z) \times   f^m_{u}(\rho,z)$.
The calibration map's correction for each point is labeled $f_u^m(\rho,z)$.
Independent uniformity calibration maps $f^m_{u}(\rho,z)$
are created for data and MC, such 
that the uniformity calibration serves to minimize any possible 
difference in position dependence of the data with respect to MC.
The capture peak on H (2.223 MeV) of neutrons from
spallation and antineutrino interactions provides a precise and copious 
calibration source to characterize the response non-uniformity over the 
full volume (both NT and GC).
The calibration map for data is shown in Figure~\ref{fig:map}.
A similar map was measured and applied to MC.
A 2D-interpolation method was developed to provide a smooth application 
of the calibration map at any point $(\rho,z)$.
%
%
%
%
%
%

\begin{figure}
  \includegraphics[scale=0.45]{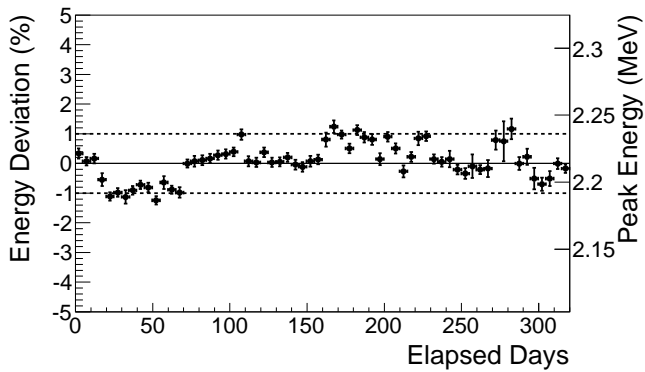}
  \caption{
    Stability of the reconstructed energy as sampled by the evolution 
in response of the spallation neutron H-capture after stability 
calibration.
    The observed steps correspond to power-cycle periods.
    The systematic uncertainty on the energy stability 
is estimated at 0.61\%.
    \label{fig:stability}}
\end{figure}

The detector response stability was found to vary in time due to two 
effects, which are accounted for and corrected by the term $f_s^m(t)$.
First, the detector response can change due to variations in readout gain 
or scintillator response.
This effect has been measured as a 
$+2.2$\% monotonic increase over 1 year using 
the response of the spallation neutrons capturing on Gd within 
the NT, shown in Figure~\ref{fig:gstability}.
Second, a few readout channels varying over time
are excluded from the calorimetry 
sum, and the average overall response decreases by $0.3$\% per channel 
excluded.
The MC is stable, so this correction is applied only to data.
The stability calibration is relative to a specific reference
time $t_{0}$.
Therefore, any response $P\!E_{\odot}(t)$ is converted to the equivalent 
response at $t_{0}$, as $P\!E^{m}_{\odot t_{0}} = P\!E^{m}_{\odot}(t) \times 
f^m_s(t)$.  
The $t_0$ was defined as the day of the first Cf source
deployment, during August 2011.
The remaining instability after calibration is shown in 
Figure~\ref{fig:stability}, as sampled with H-capture from spallation 
neutrons, and is used for the stability systematic uncertainty 
estimation.
%
%
%
%

The number $P\!E_{\odot t_0}$ per MeV is
determined by an absolute energy calibration
independently, for the data and MC.
The response in $P\!E_{\odot t_0}$ 
for H-capture 
as deployed in the center of the NT is used for the absolute
energy scale.
The absolute energy scales are found to be
$229.9~P\!E_{\odot t_0}$/MeV and $227.7~P\!E_{\odot 
t_0}$/MeV, respectively, for the data and MC,
demonstrating agreement within 1\% prior to this calibration
stage.

Discrepancies in response between the MC and data, after 
calibration, are used to estimate these uncertainties within the 
prompt energy range and the NT volume.
Table~\ref{tab:syst} summarizes the systematic uncertainty in terms of 
the remaining non-uniformity, instability and non-linearity.
\begin{table}
  \caption{Energy scale systematic errors.\label{tab:syst}}
  \begin{tabular} { | c | c  | }
    \hline
                            & Error (\%)  \\
    \hline
    Relative Non-Uniformity & 0.43 \\
    \hline
    Relative Instability    & 0.61 \\
    \hline
    Relative Non-Linearity  & 0.85 \\
    \hline
    Total                   & 1.13 \\
    \hline
  \end{tabular}
\end{table}
%
%
The relative non-uniformity systematic uncertainty was 
estimated from the calibration maps using neutrons capturing on Gd, after 
full calibration.
The rms deviation of the relative difference between 
the data and MC calibration maps is 
used as the estimator of the non-uniformity systematic uncertainty, 
and is $0.43$\%.
This result is consistent with the
analysis of all calibration sources along the $z$-axis (NT) and GT (GC).
%
%
The relative instability systematic error, discussed above, is
$0.61$\%.
%
Responses are equalized at $2.223$~MeV, but small
data/MC discrepancies in the absolute energy scale can still arise from the 
relative non-linearity across the prompt energy spectrum.
This possibility was explored by using all calibration sources in
the energy range 0.7 -- 8~MeV with deployments along the $z$-axis and GT.
Some relative non-linearity was observed ($<0.2$\%/MeV) but the pattern 
diminished when integrated over the full volume.
A $0.85$\% variation consistent with this non-linearity was measured
with the $z$-axis calibration system, and this is used as the
systematic error for relative non-linearity in Table~\ref{tab:syst}.
Consistent results were obtained when sampling with the same sources 
along the GT.

\section{Neutrino Data Analysis}
\label{sec:analysis}

\subsection{$\bar{\nu}_e$ Candidate Selection}
\label{sec:neutrino}

\begin{figure}[htb]
\begin{center}
\includegraphics[scale=0.38]{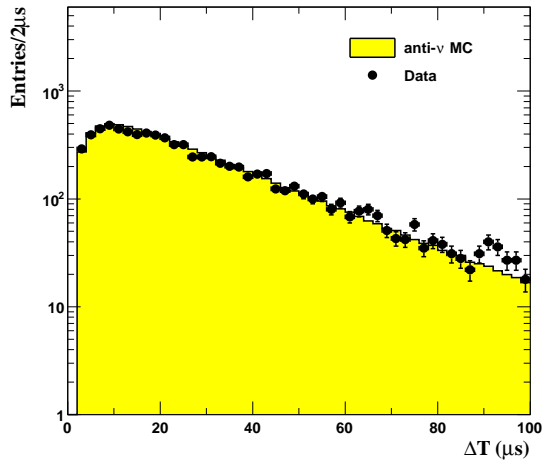}
\parbox[h]{7cm}{
\caption{Time difference between prompt and delayed triggers. Black 
dots and solid histogram show data and MC results, respectively.
\label{fig:tcut}}}
\end{center}
\end{figure}

\begin{figure}[htb]
\begin{center}
\includegraphics[scale=0.4]{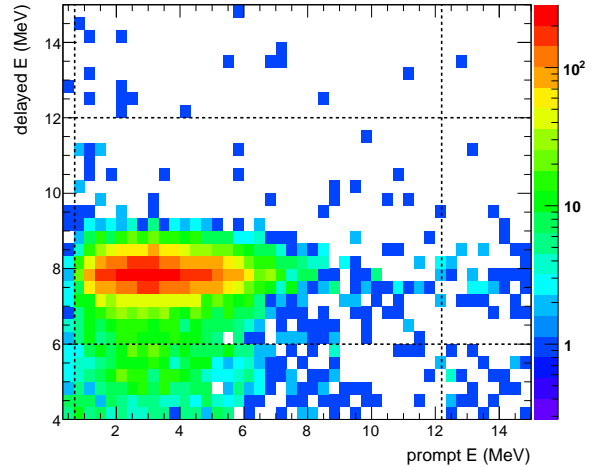}
\parbox[h]{7cm}{
\caption{Delayed energy versus prompt energy for time-correlated 
triggers. Vertical and horizontal dashed lines show the cuts applied for 
$\bar{\nu}_e$ candidates selection.
\label{fig:ecut}}}
\end{center}
\end{figure}

The $\bar{\nu}_e$ candidate selection procedure starts
in a similar way as \cite{bib:dc1}.
Events with an energy below 0.5 MeV, where 
the trigger efficiency is not 100\%, or identified as light noise 
($Q_{\rm max}/Q_{\rm tot}$ $> $ 0.09 or rms($t_{\rm start}$) $>$ $40$ ns) are 
discarded. Triggers within a 1 ms window following a 
tagged muon
are also rejected 
(see Section~\ref{sec:muon}),
in order to reduce the correlated and cosmogenic backgrounds.
The effective veto time is 4.4\% of the total run time.
Defining $\Delta T \equiv t_{\rm delayed} - t_{\rm prompt}$,
further selection consists of 4 cuts: 
\begin{enumerate}
\item time difference 
between consecutive triggers (prompt and delayed): 
$2~\mu$s $<$ $\Delta$T $<$ $100~\mu$s, as shown in 
Figure~\ref{fig:tcut}, where the lower cut 
reduces correlated backgrounds and the upper cut is determined by 
the approximately $30~\mu$s capture time on Gd; 
\item prompt 
trigger: $0.7~$MeV $ < E_{\rm prompt} < 12.2~$MeV, as illustrated 
in Figure~\ref{fig:ecut}; 
\item delayed trigger: 
$6.0~$MeV $ < E_{\rm delayed} < 12.0~$MeV (Figure~\ref{fig:ecut}) 
and $Q_{\rm max}/Q_{\rm tot}$ $<$0.055; 
\item multiplicity: no 
additional triggers from $100~\mu$s preceding the prompt 
signal to $400~\mu$s after it, with the goal of reducing 
the correlated background.   
\end{enumerate}
The IBD efficiencies for these cuts are listed in Table~\ref{tab:eff}.
\begin{table}[ht!]
\begin{center}
\begin{tabular}{| c  | c |}
\hline   
\textbf{Cut} & \textbf{Efficiency \%}\\ \hline
$E_{\rm prompt}$ & 100.0 $\pm$ 0.0 \\
$E_{\rm delayed}$ & 94.1 $\pm$ 0.6 \\
$\Delta~T$ & 96.2 $\pm$ 0.5 \\
Multiplicity & 99.5 $\pm$ 0.0 \\
Muon veto & 90.8 $\pm$ 0.0 \\
Outer Veto & 99.9 $\pm$ 0.0 \\
\hline
\end{tabular}
\caption{Cuts used in the event selection and their
efficiency for IBD events.  The OV was working for
the last 68.9\% of the data.
}
\label{tab:eff}
\end{center}
\end{table}

A preliminary sample of 9021 candidates is obtained by applying 
selections (1--4). In order to reduce the background contamination 
in the sample, candidates are rejected according to two extra 
cuts not used in \cite{bib:dc1}. First, candidates 
within a 0.5 s window after a high 
energy muon crossing the ID ($E_{\mu}> 600$~MeV) are tagged as 
cosmogenic isotope events and rejected, increasing the effective veto time
to 9.2\%. Second, candidates whose prompt 
signal is coincident with an OV trigger are also excluded as 
correlated background. Applying the above vetoes yields 8249 
candidates or a rate of 36.2 $\pm$ 0.4 events/day, uniformly distributed 
within the target, for an analysis livetime of 227.93 days. 
This rate is lower than the one presented in \cite{bib:dc1} 
due to a longer data taking period with one reactor being 
off, as well as to the new cuts reducing the background 
contamination.
Following the same selection procedure on the $\bar{\nu}_e$ MC 
sample yields 8439.6 expected events in the
absence of oscillation. 
\subsection{Accidental Background}
\label{sec:accidental}

The main source of accidental coincidences is the random
association of a prompt trigger from natural radioactivity
and a later neutron-like candidate. This background
is estimated by applying the neutrino selection
cuts described in Section~\ref{sec:neutrino} but using coincidence windows
shifted by 1 s in order to remove correlations in the time scale of 
n-captures in H and Gd.
The statistics of the sample is enhanced by using
198 windows each shifted from the previous one by 500
$\mu$s. The radioactivity rate between 0.7 and 12.2 MeV is
8.2 s$^{-1}$, while the singles rate in 6 - 12 MeV energy region
is 18 h$^{-1}$. Both rates are quite stable along the data
taking period. Finally, the accidental background rate
is found to be 0.261 $\pm$ 0.002 events per day. 
The reproducibility of our result and any possible systematic effect
are studied by repeating the procedure 30 times, i.e., taking 30 times 
198 consecutive time windows. The dispersion of these 30 measurements is 
consistent with only statistical error, so, no systematic deviation is 
found. 

\begin{figure}
\includegraphics[scale=0.4]{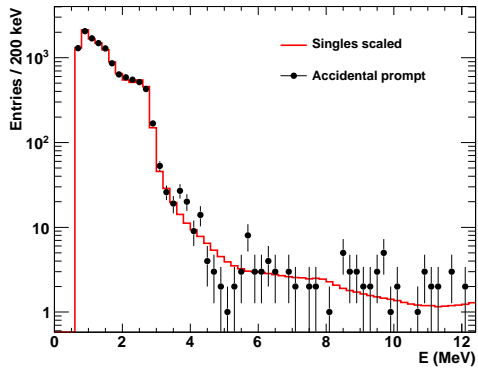}
\caption{
The accidental prompt spectrum (black circles) 
superimposed to the radioactivity energy distribution measured in Double 
Chooz scaled to the same number of entries (red line).
\label{fig:accidental}}
\end{figure}

Figure~\ref{fig:accidental} shows the accidental prompt spectrum and the 
energy distribution 
for natural radioactivity scaled to the number of accidental events;
 the agreement is excellent. The distribution is peaked at low 
energies below 3 MeV.  
The remaining light noise is included in the accidental background 
sample. Using the correlation between both 
variables $Q_{\rm max}$/$Q_{\rm tot}$ and 
rms($t_{start}$), its contribution to the accidental sample 
is estimated to be lower than 1$\%$. 

\subsection{Cosmogenic Isotopes Background}
\label{sec:cosmogenics}
The radioisotopes $^8$He and $^9$Li are products of 
spallation processes on $^{12}$C induced by cosmic 
muons crossing the scintillator volume. The $\beta n$-decays 
of these isotopes constitute a background for 
the antineutrino search. $\beta n$-emitters can be identified 
from the time- and space- correlation to their 
parent muon. Due to their relatively long 
lifetimes ($^9$Li:  $\tau=257\,$ms, $^8$He: $\tau=172\,$ms), an 
event-by-event discrimination is not possible. 
For the muon rates in our detector,
vetoing for several isotope lifetimes after each 
muon would lead to an unacceptably large loss 
in exposure. Instead, the rate is determined by an 
exponential fit to the $\Delta t_{\mu\nu} \equiv t_\mu - t_\nu$ 
profile of all possible muon-IBD candidate-pairs. 

The analysis is performed for three visible energy 
$E_\mu^{vis}$ ranges that characterize subsamples of 
parent muons by their energy deposition, not corrected
for energy non-linearities, in the ID:
\begin{enumerate}
\item Showering muons crossing the target 
value are selected by $E_\mu^{vis}>600$\,MeV and 
feature an increased probability to produce 
cosmogenic isotopes. The $\Delta t_{\mu\nu}$-fit returns a 
precise result of $0.95\pm0.11$ events/day 
for the $\beta n$-emitter rate.
\item In the $E_\mu^{vis}$ range from 
275 to 600\,MeV, muons crossing GC and target still give a 
sizable contribution to isotope production 
of $1.08\,\pm\,0.44$ events/day. To 
obtain this result from a $\Delta t_{\mu\nu}$ 
fit, the sample of muon-IBD pairs has to be cleaned by a 
spatial cut on the distance of closest approach from the
muon to the IBD candidate of $d_{\mu\nu}<80$\,cm 
to remove the majority of uncorrelated 
pairs. The corresponding cut efficiency is 
determined from the lateral distance profile obtained for 
$E_\mu^{vis}>600\,$MeV. The approach is validated 
by a comparative study of cosmic neutrons that show 
an almost congruent profile with very little 
dependence on $E_\mu^{vis}$ above 275\,MeV.
\item  The Cut $E_\mu^{vis}<275$\,MeV selects muons crossing 
only the buffer volume or the rim of the GC. For 
this sample, no production of $\beta n$-emitters 
inside the target volume is observed. An upper limit of 
$<0.3$ events/day can be established based on a 
$\Delta t_{\mu\nu}$ fit for $d_{\mu\nu}<80$\,cm. Again, 
the lateral distribution of cosmic neutrons has 
been used for determining the cut efficiency. 
\end{enumerate}
\par 
The overall rate of $\beta n$-decays found  is 
$2.05^{+0.62}_{-0.52}$ events/day. The result of a 
similar analysis based on the IV 
muon tracking agrees within the uncertainty.

Accidental coincidences containing the 
$\beta$-decay of the isotope  {$^{12}$B} either as prompt or as 
delayed event feature a time correlation to 
the parent muons producing the {$^{12}$B}. In the $\Delta 
t_{\mu\nu}$ profile, these events are represented 
by a decay function with $\tau({^{12}B})=29$\,ms. 
However, these events were removed very efficiently 
from the data set used for $^9$Li analysis by 
imposing a maximum distance cut of 90\,cm between 
prompt and 
delayed events, introducing a negligible inefficiency of $\sim$1\%.

The correlation of a cosmogenic isotope to the 
showering muons has been exploited
to impose a partial veto of this background for 
the final fit analysis. Vetoing all IBD candidates within 0.5\,s 
following a muon of $E_\mu^{vis}>600$\,MeV, $0.89\pm0.10$ 
events d$^{-1}$ of $\beta n$-decays are 
removed from the data sample.
The  
residual cosmogenic isotope
background rate has been determined to $1.25\pm0.54$ events/day.

\begin{figure}
\includegraphics[scale=0.45]{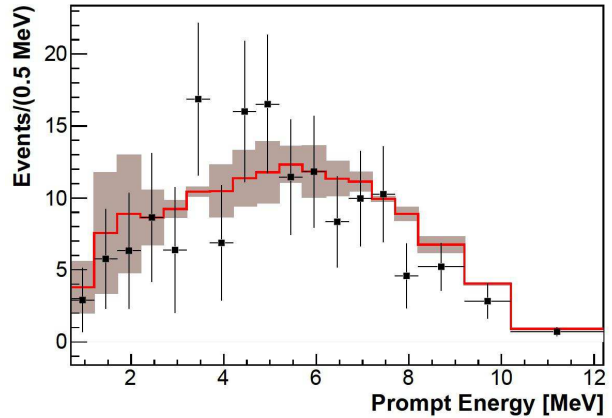}
\caption{The prompt $\beta$ spectrum of the $\beta n$-emitters
$^8$He and $^9$Li from data (black squares) and Monte Carlo
(red line),
assuming $^9$Li is the dominant contribution.
}\label{fig::lispectrum}
\end{figure}

Finally, the correlation of parent muons and 
$\beta n$-emitters has been used to extract the prompt 
$\beta$ spectrum from the data. 
Figure\,\ref{fig::lispectrum} shows a sample 
spectrum obtained for 
$E_{\mu}>620$\,MeV, a distance cut of 0.7\,m 
and a $\Delta t_{\mu\nu}$ cut of 600\,ms. The 
contamination of the sample by random coincidences 
has been statistically subtracted. Good
agreement is found for the MC spectrum used in 
the final fit analysis.

\subsection{Fast Neutrons and Stopping Muon Background}
\label{sec:fastn}

Most correlated backgrounds are rejected by the 1 ms veto time after each
tagged muon.  The remaining events arise from cosmogenic events whose parent
muon either misses the detector or deposits an energy low enough to
escape the muon tagging. Two contributions have been found: fast
neutrons (FN) and stopping muons (SM).

FN are created by muons in the inactive regions surrounding
the detector. Their large interaction length allows them to cross the
detector and capture in the ID, causing both a prompt trigger by
recoil protons and a delayed trigger by capture on Gd. An
approximately flat prompt energy spectrum is expected; a slope could
be introduced by acceptance and scintillator quenching
effects. The time and spatial correlation
distribution of FN are indistinguishable from those of \anu events.

The selected SM arise from muons entering through the chimney,
stopping in the top of the ID, and eventually decaying. The short muon
track mimics the prompt event, and the decay Michel electron mimics
the delayed event. SM candidates are localized in space in the top of
the ID under the chimney, and have a prompt-delayed time distribution
following the $2.2~\mu\text{s}$ muon lifetime.

The correlated background has been studied by extending the selection
on $E_{\text{prompt}}$ up to 30~MeV. No IBD events are expected in the
interval $12~\text{MeV} \leq E_{\text{prompt}} \leq 30~\text{MeV}$. FN
and SM candidates were separated via their different correlation time
distributions. A 
$97^{+3}_{-8}$\%
pure sample of FN is obtained for
$\Delta T > 10~\mu\text{s}$, and a $(88\pm 7)\%$ pure sample of SM is
obtained for $\Delta T < 10~\mu\text{s}$. These samples of FN and SM
can be used to estimate their rate. The observed prompt energy
spectrum is consistent with a flat continuum between 12 and 30 MeV,
which extrapolated to the IBD selection window provides a first
estimation of the correlated background rate of $\approx
0.75$~events/day.  The accuracy of this estimate depends on the
validity of the extrapolation of the spectral shape.  Next we describe
a measurement of the FN and SM spectral shapes including the IBD region,
obtained by using the IV and OV to tag samples of FN and SM.

The DAQ reads out the IV upon any ID trigger, lowering the IV
detection threshold to $\sim 1~\text{MeV}$,
and making the IV sensitive 
to FN via the detection of proton recoils and captures on H. The 
IV-tagging is implemented by demanding at least 2 IV PMT hits 
leading to ($33~\pm~5$)\% tagging efficiency with no contribution 
by single PMT energy depositions.
There is a very low probability of accidental IV tagging due to 
any IV energy deposition in the 256 ns coincident
readout window.

The OV-tagging, when available, is especially sensitive to 
SMs since the muon is often
detected. $(41\pm 23)\%$ of the FN and SM candidates in the
12-to-30-MeV window are tagged by the OV, of which $(74\pm 12)\%$ are
SM. OV-tagging has an accidental rate $=~0.06\%$ of the neutrino
sample and can be used to veto
events caused by muons. 

Several FN and SM analyses were performed using different
combinations of IV and OV tagging. The main analysis for the
FN estimation relies on 
IV-tagging of the prompt triggers with OV veto applied for
the IBD selection. Two sources
of backgrounds on the tagged FN sample were identified and
rejected. The first source is the combination of natural radioactivity
in the IV in an accidental coincidence with a genuine IBD, and was
reduced to 12\% by imposing a time coincidence between the ID and IV
energy depositions. The second source, a Compton scattering in both
the IV and ID in an accidental coincidence with a Gd-capture, was
reduced to 2\% by imposing a cut on the spatial distance between the
prompt and delayed candidate in the ID.  The purity of the IV-tagged
FN sample was 86\%. The remaining background was measured in an
off-time window and subtracted, thus minimizing distortions to the
energy spectrum. The FN spectral shape was found to be in agreement
with a linear model with a small positive slope. The measured
total FN rate was $(0.30\pm 0.14)$~events/day, including systematic
uncertainties from the $\Delta T$-based FN-SM separation, the
IV-tagging efficiency, and background subtractions.

Since there is no correlation between the SM prompt energy and the
delayed energy deposit of the Michel electron, a pure sample of SM was
obtained by selecting $20~\text{MeV} \leq E_{\text{delayed}} \leq
60~\text{MeV}$. The spectral shape of SM prompt energy was found to be
in agreement with a linear model with a small negative slope.
The total SM rate was measured to be $(0.34\pm
0.18)$~events/day, including systematic uncertainties.

\begin{figure}
\includegraphics[scale=0.42]{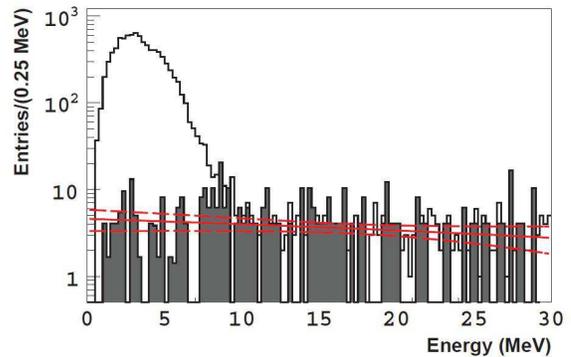}
\caption{FN and SM combined spectral model best fit 
(solid red) with $\pm\,1\sigma$ (dashed red), energy
distribution of tagged FN and SM population (gray histogram) and IBD
spectrum.
\label{fig:correlated}}
\end{figure}

Since the spectral shapes for both FN and SM are linear, a combined
analysis was performed to obtain the total spectrum shown in
Fig.~\ref{fig:correlated} and the total rate estimation $(0.67\pm
0.20)$~events/day summarized in Table~\ref{tab:data_pred}.  Consistent
results were obtained from different analysis techniques, which
included IV- and OV-tagging without OV-vetoing. The OV veto reduces
the rate of correlated backgrounds by about 30\%.

\subsection{Background Measurements}

There are four ways that can be utilized to estimate backgrounds. 
Each independent background component can be measured
by isolating samples and subtracting 
possible correlations. This is described for each component
in Sections \ref{sec:lightnoise}, \ref{sec:cosmogenics}
and \ref{sec:fastn}.
Second, we can measure each independent background component 
including spectral information when fitting for \qeq
~oscillations as is done in Section~\ref{sec:finalfit}. Third, the
total background rate is measured by comparing the 
observed and expected rates as a function of reactor power. 
Fourth, we can use the 
both-reactor-off data to measure both the rate and spectrum. 

\begin{figure}[htb]
\begin{center}
\includegraphics[scale=0.4]{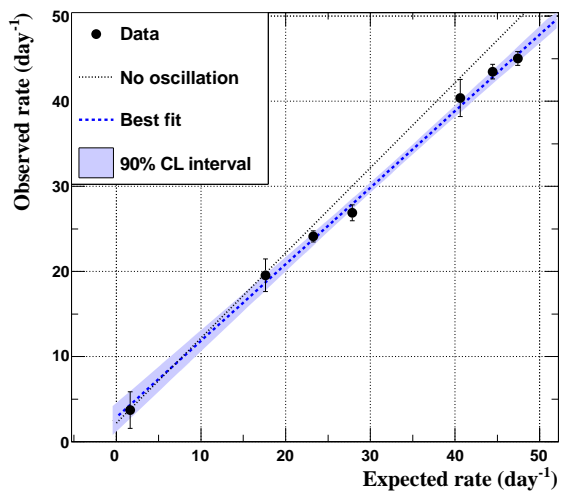}
\parbox[h]{7cm}{
\caption{Daily number of  $\bar{\nu}_e$ candidates as a function of the
expected number of  $\bar{\nu}_e$. The dashed line shows the fit to the data,
along with the 90\% C.L. band. The dotted line shows the 
expectation in the no-oscillation scenario.
\label{fig:rate}}}
\end{center}
\end{figure}

The latter two methods are used currently as cross-checks 
for the background measurements due to low statistics and 
are described here. The measured daily rate of IBD candidates 
as a function of the no-oscillation expected rate for different 
reactor power conditions is shown in Figure~\ref{fig:rate}.
The extrapolation 
to zero reactor power of the fit to the data yields 2.9 $\pm$ 1.1 events 
per day, in excellent agreement with our background estimate.
The overall rate of correlated background events that pass the IBD cuts
 is independently verified by analyzing 22.5 hours of both-reactors-off 
data. The expected neutrino signal is $<0.3$ residual 
$\bar{\nu}_e$ events. 
Three events passed the first 4 cuts in Section~\ref{sec:neutrino}.
Two events with prompt energies of 4.8 MeV and 9.4 MeV were 
associated within 30 cm and 240 ms with the closest energetic muon, 
and are thus likely to be associated with $^9$Li. Indeed, the second 
candidate is rejected by the showering muon veto. The third candidate 
at a prompt energy of 0.8 MeV features 3.5 m distance between 
prompt and delayed events and is therefore most likely a 
random coincidence. Immediately following the data set used in this 
paper, we obtained a larger data set with both-reactors-off. That will 
be the subject of a separate paper \cite{bib:offoff}.

\subsection{Neutron Detection Efficiency}
\label{sec:efficiency}
Calibration data taken with the $^{252}$Cf source were used to check the Monte Carlo prediction
for any biases in the neutron selection criteria and 
estimate their contributions to the systematic uncertainty.
\par 
The fraction of neutron captures on gadolinium is evaluated to be 86.5\%
near the center of the target, 1.5\% lower than the fraction 
predicted by simulation.
Therefore the Monte Carlo simulation for the prediction 
of the number of $\bar{\nu}_e$  events
is reduced by factor of 0.985.
After the prediction of the fraction of neutron 
captures on gadolinium is scaled to the data,
the prediction reproduces the data to 
within 0.3\% under variation of selection criteria. 
\par
\begin{figure}
\includegraphics[scale=0.5]{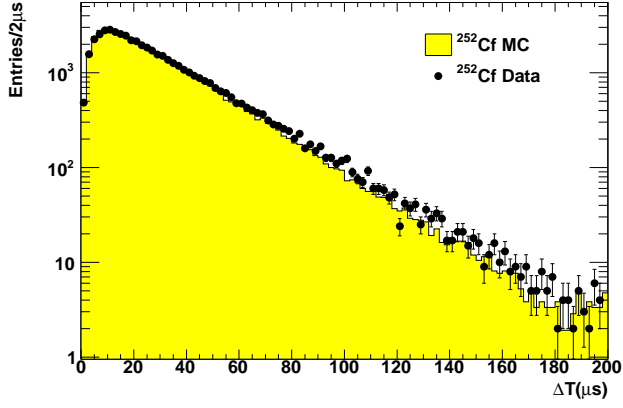}
\caption{
Time difference between prompt and delayed events
with $^{252}$Cf at the detector center.
The prompt time is determined by 7-30~MeV gamma ray.
\label{fig:californium}}
\end{figure}

 The $^{252}$Cf is also used to check the neutron capture time,
$\Delta T$.
The time difference between the prompt event and neutron
capture signal for the californium calibration data is shown
in Figure~\ref{fig:californium}.
The simulation reproduces the efficiency (96.2\%) of the 
$\Delta t_{e^+n}$ cut with an uncertainty
of 0.5\% augmented with sources deployed through the NT and GC.
\par The efficiency for Gd capture events with visible energy greater 
than 4 MeV to pass the 6 MeV cut
is estimated to be 94.1\%.
Averaged over the NT, the fraction of neutron captures on Gd accepted by the
6.0 MeV cut is in agreement with calibration data to within 0.7\%.
\par The Monte Carlo simulation indicates that the number of IBD events occurring in the
GC with the neutron captured in the NT (spill-in) slightly exceeds the number
of events occurring in the target with the neutron escaping to the gamma catcher (spill-out), by
1.35\%  $\pm$ 0.04\%(stat) $\pm$ 0.30\%(sys).
The spill-in/out effect is already included in the simulation and 
therefore no correction for this is needed.
The uncertainty of 0.3\% assigned to the net
spill-in/out current was 
quantified by varying the parameters 
affecting the process, such as gadolinium concentration in 
the target scintillator and hydrogen fraction
in the gamma-catcher fluid within its tolerances. Moreover 
the parameter variation was performed with multiple
 Monte Carlo models at low neutron energies.

\section{Oscillation Analysis}
\label{sec:finalfit}


The oscillation analysis is based on a combined 
fit to antineutrino 
rate and spectral shape.  IBD candidates are 
selected as described in Section~\ref{sec:neutrino}.  The 
data are compared to the Monte Carlo signal and background 
events from high-statistics 
samples.  The same selections are applied to both signal and 
background, with corrections 
made to Monte Carlo only when necessary to match detector 
performance metrics.

The oscillation analysis begins by separating the data into  
18 variably-sized bins between 0.7 and 12.2 MeV.  
Two integration periods are used in the fit to help 
separate background and signal flux.
One set contains data periods where one 
reactor is operating at less 
than 20\% of its nominal thermal power, according 
to power data provided by EDF, while the other set contains 
data from all other times, typically when both reactors are running.  
All data end up in one of the two integration periods.
Here, we denote the number of observed 
IBD candidates in each of the bins as $N_i$, where 
$i$ runs over the combined 36 
bins of both integration periods.  The use of 
multiple periods of data integration 
takes advantage of the different signal/background 
ratios in each period, as 
the signal rate varies with reactor power while the 
backgrounds remain constant in time.
This technique adds information about background behavior 
to the fit.  
The distribution of IBD candidates between the two integration 
periods is given in 
Table~\ref{tab:data_pred}.

A prediction of the observed number of signal and background 
events is constructed 
for each energy bin, following the same integration period 
division as the data:  

\begin{equation}
N^\text{pred}_i = \sum\limits_{R = 1,2}^{\text{Reactors}} 
N^{\nu,R}_i + \sum\limits_{b}^{\text{Bkgnds.}} N^{b}_{i}
\label{eqn:npred}
\end{equation}
where  $N_i^{\nu,R} = P_(\anum \rightarrow \anum) N_i^{\text{exp},R}$;
$P_{\anum \rightarrow \anum}$ is the neutrino survival 
probability from the well-known oscillation formula and 
$N_i^{\text{exp},R}$ is given by Equation~\ref{eq:prediction}.
The index $b$ runs over the three backgrounds: cosmogenic 
isotope; correlated; 
and accidental.  The index $R$ runs over the 
two reactors, Chooz B1 and B2.
 
Background populations were calculated based on 
the measured rates and the livetime 
of the detector during each integration period.  
Details on the signal prediction normalization 
can be found in Sec.~\ref{subsec:nurate}.  Predicted populations 
for both null-oscillation signal and 
backgrounds may be found in Table~\ref{tab:data_pred}.

\begin{table}[ht!]
\begin{center}
\begin{tabular}{| c | c | c | c |}
\hline   
& \textbf{Reactors} & 
\textbf{One Reactor} & \textbf{Total} \\
& \textbf{Both On} & 
\textbf{$P_{th} <$ 20\%} &  \\
Livetime [days] & 139.27 & 88.66 & 227.93 \\
IBD Candidates & 6088 & 2161 & 8249 \\
\hline \hline
$\nu$ Reactor B1 & 2910.9 & 774.6 & 3685.5 \\
$\nu$ Reactor B2 & 3422.4 & 1331.7 & 4754.1 \\
Cosmogenic Isotope & 174.1 & 110.8 & 284.9 \\
Correlated FN  \& SM & 93.3 & 59.4 & 152.7 \\
Accidentals & 36.4 & 23.1 & 59.5 \\
\hline
Total Prediction & 6637.1 & 2299.7 & 8936.8 \\
\hline
\end{tabular}
\caption{Summary of observed IBD candidates, with corresponding 
signal and background predictions for each integration period, before any 
oscillation fit results have been 
applied.}
\label{tab:data_pred}
\end{center}
\end{table}

Systematic and statistical uncertainties are 
propagated to the fit by 
the use of a covariance matrix $M_{ij}$ in 
order to properly account for 
correlations between energy bins. 
The sources of uncertainty $A$ are listed in Table~\ref{tab:systematics}.
\begin{equation}
M_{ij} = M^\text{sig.}_{ij} + M^\text{det.}_{ij} + M^\text{stat.}_{ij} 
+ M^\text{eff.}_{ij} + \sum\limits_{b}^{\text{Bkgnds.}} M^\text{b}_{ij}
\label{eqn:covmat}
\end{equation}
\noindent
Each term $M^\text{A}_{ij} = 
\text{cov}(N^\text{pred}_i,N^\text{pred}_j)_{A}$
on the right-hand side of 
Equation~\ref{eqn:covmat} represents the covariance of $N^\text{pred}_i$ 
and $N^\text{pred}_j$ due to uncertainty $A$.
The normalization uncertainty associated 
with each of the matrix contributions may be found from 
the sum of each matrix: these are summarized in 
Table~\ref{tab:systematics}.  
Many sources of uncertainty contain spectral shape 
components which do not directly 
contribute to the normalization error, but do provide 
for correlated uncertainties 
between the energy bins.  The signal covariance matrix 
$M^\text{sig.}_{ij}$ is calculated taking 
into account knowledge about the predicted neutrino 
spectra.
The $^9$Li matrix contribution contains spectral shape 
uncertainties estimated using different Monte Carlo 
event generation parameters, as 
described in Sec.~\ref{sec:monte}.  The slope of the FN/SM
spectrum is allowed to 
vary from a nearly-flat spectrum following the measurements 
described in Section~\ref{sec:fastn}.  
Since 
accidental background uncertainties 
are measured to a high precision from many 
off-time windows, they 
are included as a diagonal covariance matrix.  

The elements of the covariance matrix contributions 
are recalculated as a function of the 
oscillation and other parameters (see below) at each 
step of the minimization.  This maintains 
the fractional systematic uncertainties as the 
bin populations vary from the 
changes in the oscillation and fit parameters.

\begin{table}[ht!]
\begin{center}
\begin{tabular}{| c | c |}
\hline
\textbf{Source} & \textbf{Uncertainty [\%]} \\
\hline
Reactor Flux & 1.67\% \\
Detector Response & 0.32\% \\
Statistics & 1.06\% \\
Efficiency & 0.95\% \\
Cosmogenic Isotope Background & 1.38\% \\
FN/SM & 0.51\% \\
Accidental Background &  0.01\%   \\ 
\hline
\textbf{Total} & 2.66\%  \\ \hline
\end{tabular}
\caption{Summary of signal and background normalization 
uncertainties in this analysis relative
to the total prediction.}
\label{tab:systematics}
\end{center}
\end{table}

A fit of the binned signal and background data to a 
two-neutrino oscillation hypothesis was 
performed by minimizing a standard $\chi^2$ function:
\begin{eqnarray}
\nonumber \chi^2  &=& \sum\limits_{i,j}^{36} 
\left( N_i - N^\text{pred}_i \right) \\
& \times & \left( M_{ij} \right)^{-1} 
\left( N_j - N^\text{pred}_j \right)^{\text{T}} \nonumber \\
& &+ \frac{\left( 
\epsilon_{FN/SM}-1 \right)^2}{\sigma^{2}_{FN/SM}} 
+ \frac{\left( \epsilon_{{}^9\text{Li}} - 1 
\right)^2}{\sigma^{2}_{{}^9\text{Li}}} \nonumber \\
& &+ \frac{\left( \alpha_E - 1 \right)^2 
}{\sigma^2_{\alpha_E}}
+ \frac{\left( \dms - \left( \dms \right)_\text{MINOS} \right)^2 
}{\sigma^2_\text{MINOS}}
\label{eqn:chi2_cov}
\end{eqnarray}
\noindent
The use of energy spectrum information in this analysis 
allows additional information 
on background rates to be gained from the fit, in particular
because of the small number of IBD events between 8 and 12 MeV.  
The two fit parameters $\epsilon_{FN/SM}$ and 
$\epsilon_{{}^9\text{Li}}$ 
are allowed to vary as part of the fit, and they
scale the rates of the two 
backgrounds 
(correlated  and cosmogenic isotope).  The rate of accidentals is not 
allowed to vary since its 
initial uncertainty is precisely determined by the measurement 
method described in Sec.~\ref{sec:accidental}.  
The energy scale for predicted signal and ${}^9$Li events 
is allowed to vary linearly according to the $\alpha_E$ parameter 
with an uncertainty $\sigma_{\alpha_E} = 1.13$\%.  
A final parameter constrains the mass splitting $\dms$ using the MINOS
measurement~\cite{bib:minosdm2} of 
$\dms = (2.32 \pm 0.12) \times 10^{-3}$ eV$^{2}$, where we have symmetrized 
the error. This error includes the uncertainty introduced by relating the 
effective mass-squared difference observed in a $\nu_\mu$ disappearance 
experiment to the one relevant for reactor experiments, and the 
ambiguity due to the type of the neutrino mass hierarchy, see 
e.g.~\cite{bib:nunokawa}.
Uncertainties for these parameters, $\sigma_{FN/SM}$, 
$\sigma_{{}^9Li}$, 
and $\sigma_\text{MINOS}$, are listed as the initial values 
in Table~\ref{tab:pulls}.

The best-fit gives $\stot = 0.109 \pm 0.030 ~(\text{stat.}) 
\pm 0.025 ~(\text{syst})$ at 
$\dms = 2.32 \times 10^{-3}$ eV$^{2}$, with a $\chi^2/\text{NDF} = 
42.1/35$.  We used the MINOS measured $\dms$
value as a constraint for our \qeq ~measurement, 
but a two parameter fit without the MINOS $\dms$ in the region
$\dms~<~0.01$~eV$^2$ gives a $\dms$ value of 
2.7 $\pm$ 1.9 $\times~10^{-3}$ eV$^2$, which is fully consistent with MINOS.
The fit gives $\stot$ = 0.093 $\pm$ 0.078 which is consistent
with our fit for $\tot$ using MINOS.

Table~\ref{tab:pulls} gives the resulting values of the fit parameters 
and their uncertainties.  Comparing the values  
with the ones used as input to the fit in Table~\ref{tab:data_pred} 
we conclude 
that the background rate and uncertainties are further constrained 
in the fit, 
as well as the energy scale.

\begin{figure*}[htb!]
\centering
\includegraphics[width=0.95\textwidth]{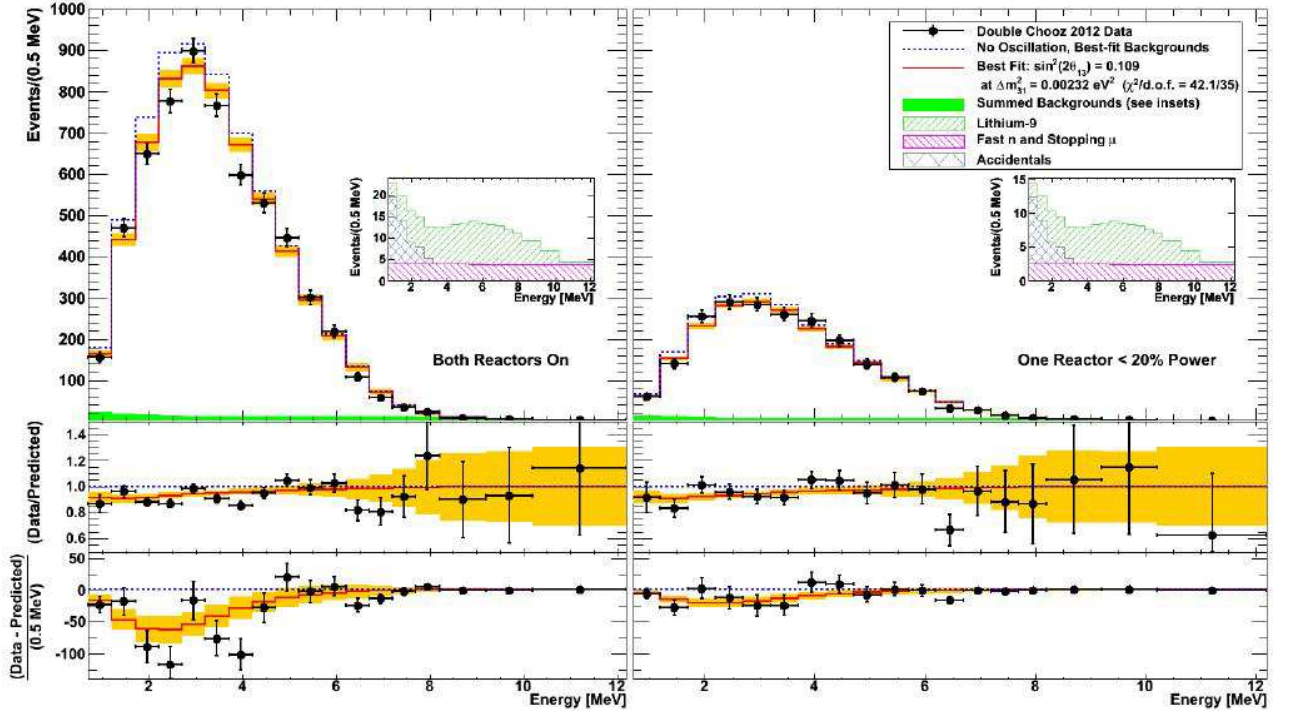}
\caption{Measured prompt energy spectrum for each integration period 
(data points) superimposed 
on the expected prompt energy spectrum, including backgrounds (green 
region), for the 
no-oscillation (blue dotted curve) and best-fit (red solid curve) at 
$\stot = 0.109$ and $\dms = 2.32 \times 10^{-3}$ eV$^2$.  Inset: 
stacked spectra of backgrounds.  
Bottom: differences between data and no-oscillation prediction 
(data points), 
and differences between best fit prediction and no-oscillation 
prediction (red curve).  The orange band represents the systematic 
uncertainties on the best-fit prediction.
\label{fig:prompt_spect}}
\end{figure*}

\begin{figure}[htb!]
\centering
\includegraphics[width=0.42\textwidth]{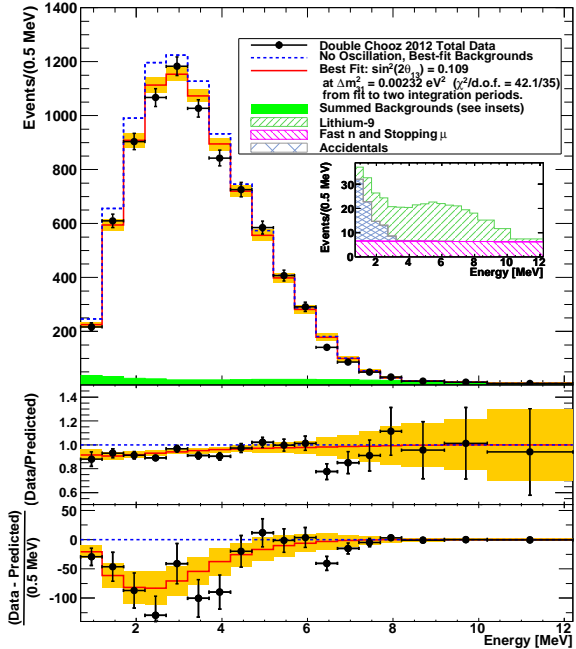}
\caption{Sum of both integration periods 
plotted in the same manner as Figure \ref{fig:prompt_spect}.
\label{fig:merged_prompt_spect}}
\end{figure}

The final measured spectrum and the best-fit spectrum are shown in 
Figure~\ref{fig:prompt_spect} for the new and old data sets, and
for both together in Figure~\ref{fig:merged_prompt_spect}.  

\begin{table}[ht!]
\begin{center}
\begin{tabular}{| c | c | c |}
\hline
\textbf{Fit Parameter} & \textbf{Initial Value} & \textbf{Best-Fit Value} 
\\
\hline
$^9$Li Bkg. $\epsilon_{{}^9\text{Li}}$ & (1.25 $\pm$ 0.54) 
d$^{-1}$ & (1.00 $\pm$ 0.29) d$^{-1}$ \\
FN/SM Bkg. $\epsilon_{FN/SM}$ & (0.67 $\pm$ 
0.20) d$^{-1}$ & (0.64 $\pm$ 0.13) d$^{-1}$ \\
Energy Scale $\alpha_E$ & 1.000 $\pm$ 0.011  & 0.986 $\pm$ 0.007 \\
$\dms$ (10$^{-3}$ eV$^2$)  & 2.32 $\pm$ 0.12 & 2.32 
$\pm$ 0.12 \\
\hline
\end{tabular}
\caption{Parameters in the oscillation fit.  Initial values 
are determined 
by measurements of background rates or detector calibration data.  
Best-fit values are outputs of the minimization procedure.}
\label{tab:pulls}
\end{center}
\end{table}

An analysis comparing only the total observed number of IBD candidates 
in each integration period 
to the expectations produces a best-fit of $\stot = 0.170 \pm 0.052$ at 
$\chi^2/\text{NDF} = 0.50/1$.  
The compatibility probability for the rate-only and rate+shape 
measurements is about 30\%
depending on how the correlated errors are handled between the two 
measurements.

\begin{figure}[htbp!]
  \centering
  \subfigure[First Publication Data Set]{\includegraphics[width=0.42
\textwidth]{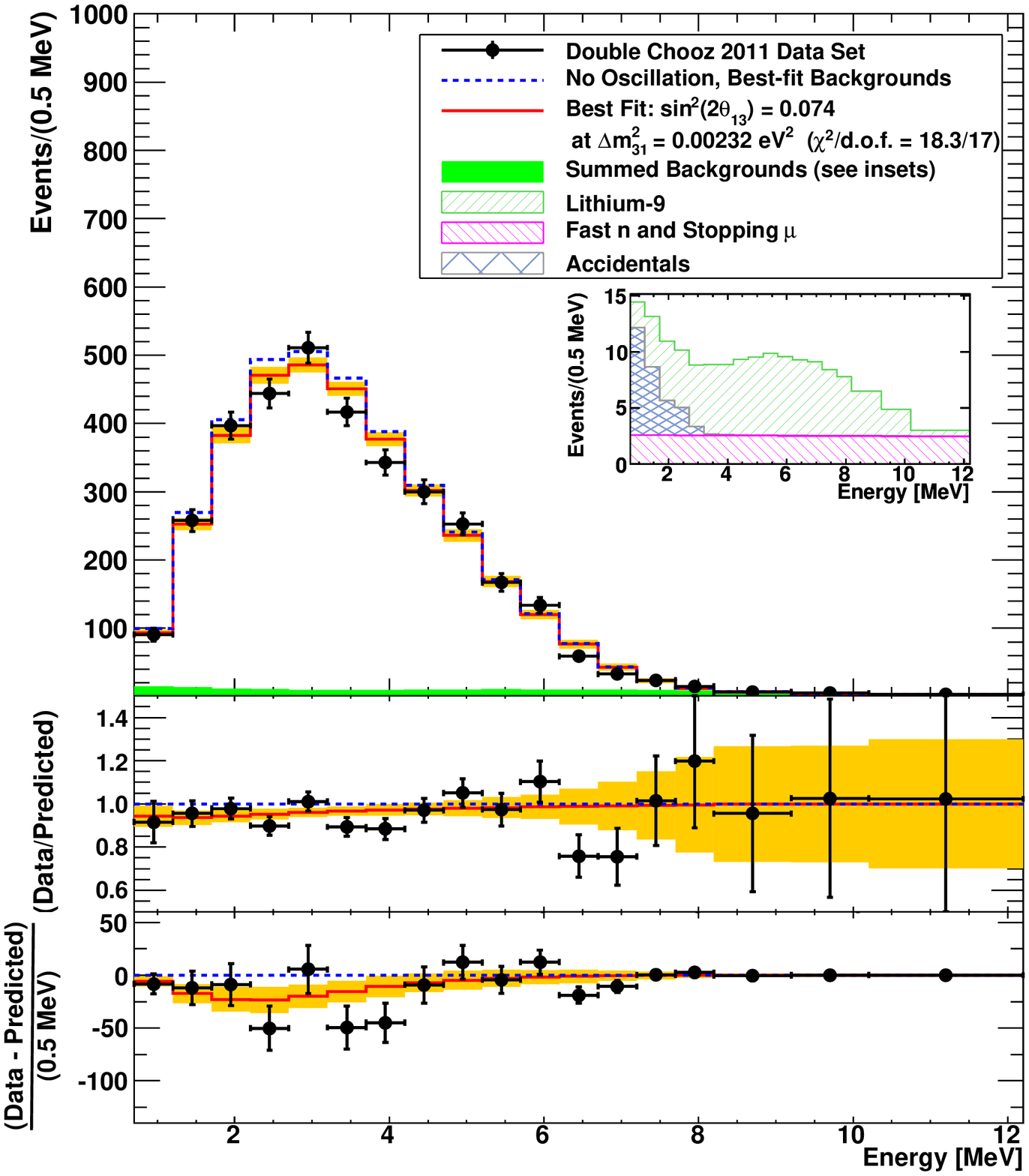}}
  \subfigure[Data Since First Publication]{\includegraphics[width=0.42
\textwidth]{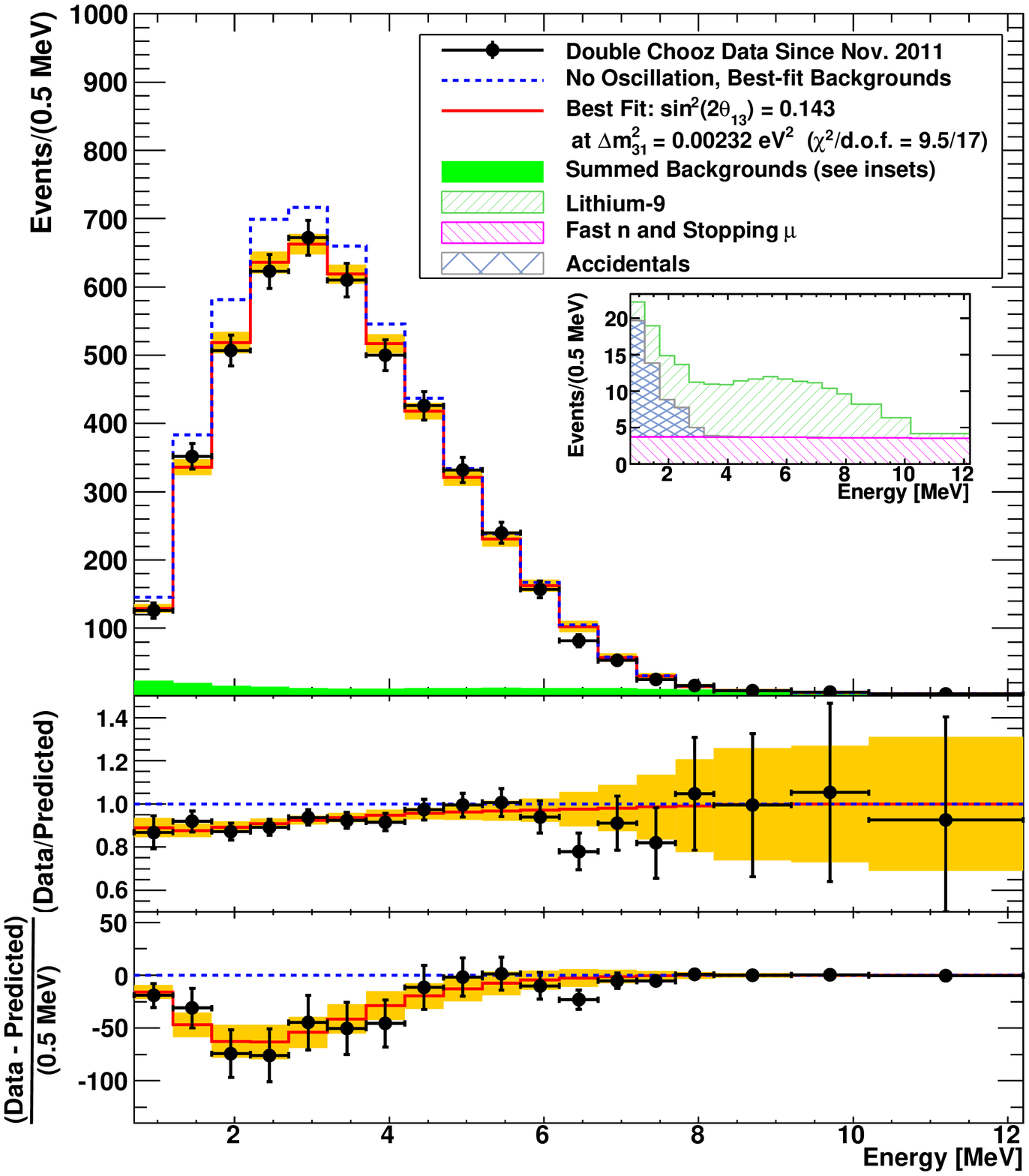}}
  \caption{Data and best-fit spectrum from 
applying current analysis 
techniques 
    to the data set used to produce the first Double Chooz 
publication (a), and 
    data taken since that publication (b), plotted in the
same manner as Figure~\ref{fig:prompt_spect}.
}
  \label{fig:dc1p}
\end{figure}

A re-processing of the data set used for the first Double Chooz 
publication \cite{bib:dc1}
was performed using the current analysis techniques. A fit using only a 
single 
integration period yielded a best-fit value of $\stot = 0.0744 ~\pm~ 
0.046$ 
with 
$\chi^2/\text{NDF} = 18.3/17$.  An analysis of only the data taken since 
the 
first publication yielded a best-fit of $\stot = ~0.143~ \pm 0.043$ with 
$\chi^2/\text{NDF} = 9.54/17$.  The data and best-fit spectra for each of 
these 
cases is shown in Figure~\ref{fig:dc1p}.

Our predicted fission cross section is 5.723 $\pm$ 0.096 $\times 10^{-43}$ 
cm$^2$/fission using the Bugey4 anchoring measurement and corresponding to 
the values of $\alpha_k$ in Table~\ref{tab:eperfiss}.  The background 
subtracted reactor antineutrino event rate is 7751.9 events, corresponding 
to 91.85\% of the rate expected in the absence of oscillations.  Our 
measured fission cross section is 5.257 
$\pm~0.056 ~(\text{stat.})~\pm~0.105 ~(\text{syst})$ $\times 10^{-43}$
cm$^2$/fission.

A further cross-check of the analysis was carried out by imposing cuts to 
eliminate the vast majority of the cosmogenic isotope background at
the cost of reduced livetime.  
The best-fit case of this analysis was found at $\stot = 0.109 
\pm 0.044$ and 
$\dms = 2.32 \times 10^{-3}$ eV$^2$, in good agreement with the 
standard analysis.


Confidence intervals for the 
standard analysis
were determined using 
a frequentist 
technique~\cite{bib:feldman}. 
This approach accommodates the fact that the true $\chi^2$ 
distributions may not be Gaussian and is useful for 
calculating the probability of excluding the 
no-oscillation hypothesis.
This study 
compared the data to 10,000 simulations generated at each of 21 
test points in the range 
$0 \leq \stot \leq 0.25$.  A $\Delta\chi^2$ statistic, 
equal to the difference 
between the $\chi^2$ at the test point and the $\chi^2$ at the best 
fit, was used to 
determine the region in $\stot$ where the 
$\Delta\chi^2$ of the data 
was within the given confidence probability.  The allowed 
region at 68\% (90\%) CL 
is 0.067 (0.043) $<$ $\stot$ $<$ 0.15 (0.18).  
An analogous 
technique shows that the data excludes the no-oscillation 
hypothesis at 99.8\% ($2.9 \sigma$).


\section{Conclusion}
\begin{figure}[htbp!]
\centering
\includegraphics[width=0.5\textwidth]{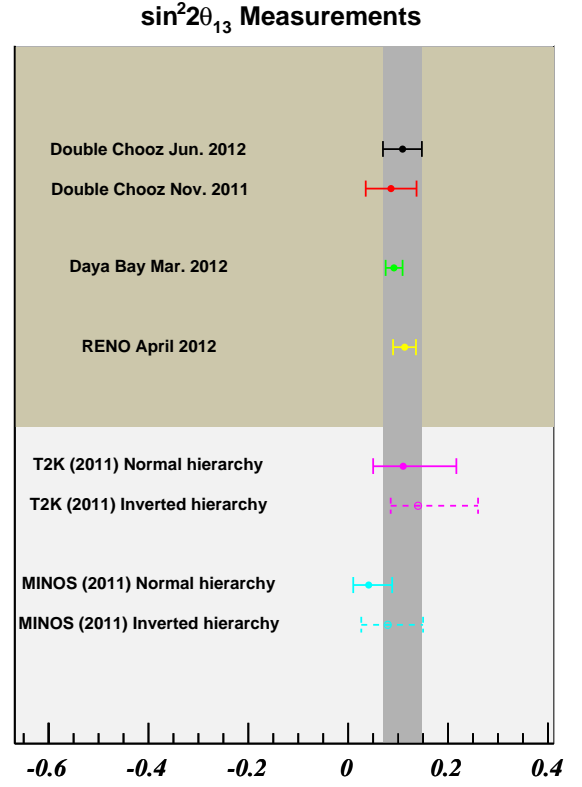}
\caption{Comparison of recent reactor- and accelerator-based 
measurements of $\stot$ from 
this analysis, the first Double Chooz publication~\cite{bib:dc1}, 
Daya Bay~\cite{bib:db1}, RENO~\cite{bib:reno}, 
T2K~\cite{bib:t2k}, and MINOS~\cite{bib:minose}.  
Error bars correspond to $1\sigma$.
For 
T2K and MINOS the CP phase $\delta$ has been 
fixed (arbitrarily) to $\delta = 0$.
\label{fig:other_expts}}
\end{figure}

A comparison of this analysis result to other recent $\stot$ measurements  
by other experiments is shown in Figure~\ref{fig:other_expts}.  The
figure shows published results, though we note that 
new results from
Daya Bay, MINOS and T2K have been shown at conferences
but are not yet published \cite{bib:nu2012}.
The values for $\stot$ from 
the various experiments are in excellent agreement with the 
results reported here.  However this result is unique in its
incorporation of energy dependence in the analysis.

Double Chooz has found evidence for a non-zero value of
\qeq ~from the rate and energy spectrum of reactor neutrino
candidates at a distance of 1050~m from two reactors.  It is
the first evidence for this parameter using the energy
spectrum from reactor neutrinos, rather than simply their
rate.
We find a best fit value and 1$\sigma$ error to be 
$\stot$ = 0.109 $\pm$ 0.030 (stat) $\pm$
0.025 (syst).  The data
is inconsistent with the assumption that
oscillations are absent with a CL of 99.8\% CL (2.9$\sigma$).

We thank the French 
electricity company EDF, the European fund FEDER, the R\'{e}gion de 
Champagne Ardenne, the D\'{e}partement des Ardennes and 
the Communaut\'{e} des Communes Rives de Meuse. We acknowledge 
the support of CEA and CNRS/IN2P3 in France, 
French LabEx UnivEarthS,
Ministry of Education, Culture, Sports, 
Science and Technology of Japan (MEXT) and Japan Society for the 
Promotion of Science (JSPS), the Department of Energy and the National Science Foundation 
of the United States, the Ministerio de Ciencia e Innovaci\'{o}n (MICINN) of 
Spain, the Max Planck Gesellschaft and the Deutsche Forschungsgemeinschaft 
DFG (SBH WI 2152), the Transregional Collaborative Research Center 
TR27, the Excellence Cluster "Origin and Structure of the Universe" and 
the Maier-Leibnitz-Laboratorium Garching, the Russian Academy of 
Science, the Kurchatov Institute and RFBR (the Russian Foundation 
for Basic Research), the Brazilian Ministry of Science, Technology and 
Innovation (MCTI), the Financiadora de Estudos e Projetos (FINEP), the 
Conselho Nacional de Desenvolvimento 
Cient\'{i}fico e Tecnol\'{o}gico (CNPq), the S\~{a}o Paulo Research 
Foundation (FAPESP), the Brazilian Network for High Energy 
Physics (RENAFAE) in Brazil and the computer center CCIN2P3.

\end{document}